\documentclass{aastex} 
\usepackage{emulateapj5,rotating}
\shorttitle{Variable Stars in Lupus}
\shortauthors{Weldrake \& Bayliss}

\begin{document}
\title{A Deep Catalog of Variable Stars in a 0.66deg$^{2}$ Lupus Field}

\author{David T. F. Weldrake} \affil{Max Planck Institut f\"ur Astronomie, K\"onigstuhl 17, D-69117, Heidelberg, Germany}
\email{weldrake@mpia-hd.mpg.de}

\author{Daniel D. R. Bayliss} \affil{Research School of Astronomy and Astrophysics, Australian National University, Mount Stromlo Observatory, Cotter Road, Weston Creek, ACT 2611, Australia}
\email{daniel@mso.anu.edu.au}

\begin{abstract}
We have conducted a wide-field photometric survey in a single 52$'$$\times$52$'$ field towards the Lupus Galactic Plane in an effort to detect transiting Hot Jupiter planets. The planet Lupus-TR-3b was identified from this work. The dataset also led to the detection of 494 field variables, all of which are new discoveries. This paper presents an overview of the project, along with the total catalog of variables, which comprises 190 eclipsing binaries (of contact, semi-contact and detached configurations), 51 miscellaneous pulsators of various types, 237 long period variables (P$\ge$2d), 11 $\delta$ Scuti stars, 4 field RR Lyrae (3 disk and 1 halo) and 1 irregular variable. Our survey provides a complete catalog of W UMa eclipsing binaries in the field to V$=$18.8, which display a Gaussian period distribution of 0.277$\pm$0.036d. Several binary systems are likely composed of equal mass M-dwarf components and others display evidence of mass transfer. We find 17 candidate blue stragglers and one binary that has the shortest period known, 0.2009d (V$=$20.9). The frequency of eclipsing binaries (all types) is found to be 1.7$\pm$0.4$\times$10$^{-3}$ per star, substantially higher (by a factor of 3-10) than previously determined in the haloes of the globular clusters 47 Tuc and $\omega$ Cen. This indicates that cluster dynamics aids mass segregation and binary destruction. 
\end{abstract}

\keywords{binaries: eclipsing --- binaries: general --- stars: variables: $\delta$ Scuti --- other}

\section{Introduction}
The discovery rate of field variable stars has dramatically increased in recent years due to the advent of large-scale photometric surveys with wide fields of view. Many such surveys are dedicated to the detection of transiting short period planets or microlensing events (for example \citet{Woz2004} and references there-in). They are perfectly suited to the detection of variable stars \citep{Sos2006,Alc2003}, as they incorporate long temporal baselines, high resolution imagery, and achieve simultaneous high precision photometry for tens of thousands of stars. 

Variable stars provide important information on the frequency, nature and evolution of stellar variability (of all  types) in various regions of the Galaxy. Binary stars provide information on stellar mass-radius relationships across many orders of magnitude, particularly for the little-understood relation for the lowest mass stars. Pulsating variables (such as $\delta$ Scuti stars) offer insights into the internal structure and evolution of main sequence and post-main sequence objects. RR Lyrae and Cepheids are often used as distance indicators and their identification (particularly in stellar clusters) provides additional distance information for comparing stellar properties with that provided by theoretical isochrones.

Unlike microlensing surveys, transit surveys have temporal resolution of several minutes, permitting the detection of variability on very short time-scales, as well as longer period variability depending on dataset properties. \citet{Kane2005} present an example of a variable star catalog from a wide angle transit survey in the general field, containing the types of variability commony found. Similarly, \citet{W2004,W2007a} present the variables identified during a transit survey of the haloes of 47 Tucanae (47 Tuc) and $\omega$ Centauri ($\omega$ Cen). Open clusters have also been targeted (\citealt{PB2006} and references there-in) and the Permanent All Sky Survey \citep{Deeg2004} has the goal of permanently tracking variable stars in the whole sky with high temporal resolution.

This paper presents the total catalog of 494 variable stars identified during a deep, wide single-field survey for transiting Hot Jupiter planets ($\sim$Jupiter-mass planets with orbital periods of a few days) towards the Lupus Galactic Plane. In addition to searching for transiting planets, this survey acts as a control field for our previous 47 Tuc and $\omega$ Cen transit surveys, both of which produced significant null results. One object, Lupus-TR-3b, that is almost certainly a transiting low-mass Hot Jupiter has been identified in our survey \citep{W2007b}. The survey also serves as an excellent test into the techniques and strategies for use in the soon-to-begin 5.7deg$^{2}$ SkyMapper transit survey \citep{BS2007}. SkyMapper has the capacity to discover dozens of new transiting planets and thousands of variable stars in the near future.

Here we present the properties of all 494 variables, and illustrative V+R lightcurves for the majority of the catalog. We detail a preliminary analysis on their occurrence rates and likely nature, particularly in comparison to previous work in the fields of the globular clusters 47 Tuc and $\omega$ Cen \citep{W2004,W2007a,W2007c}. We find a significant difference in the occurrence rate for binary stars, with our field having three times the observed binary frequency of $\omega$ Cen and ten times that of 47 Tuc. This supports mass segregation, particularly for 47 Tuc, and overall binary destruction in globular clusters due to cluster dynamics. Several of our binary systems are likely composed of M dwarf components, of interest in the study of low mass stars, and several other binaries display evidence of mass transfer, important for studies into binary system evolution and stellar interactions. We have also identified 17 blue straggler candidates and a binary which very likely contains a $\delta$ Scuti component.

Section 2 of this paper details the observational strategy and data reduction techniques. Section 3 describes the production of the stellar time-series, the removal of dataset systematics, resulting photometric accuracy, and stellar colors and astrometry.  Section 4 details the variability search methods and dataset completeness. Section 5 presents the variable catalog, detailing each main type of variable and provides the derived binary frequency in our field, with a comparison to the globular clusters and distance estimates for our four RR Lyrae stars. We conclude in Section 6.

\section{Observations and Data Reduction}
Using the Australian National University 1m telescope at Siding Spring
Observatory, a single 54$'$$\times$54$'$ Wide Field Imager (WFI) field
was observed for 53 nights, 26 contiguous nights in June 2005 and 27
nights in June 2006. The WFI detector consists of a 4$\times$2 array
of 2048$\times$4096 pixel back-illuminated CCDs, arranged to produce a
total array of 8K$\times$8K pixels. The detector scale is 0.$''$38 per pixel at the Cassegrain focus, well matched to the median seeing at the site (mean seeing of 2.2$''$) allowing a 
suitable sampling of the stellar point-spread function (PSF).  During
the course of this survey, only 7 of the 8 WFI CCDs were operational, resulting in an effective
survey field of 0.66 deg$^{2}$.

The survey field is a few degrees from the Galactic Plane and is
centered at RA$=$15$^{\rm{h}}$30$^{\rm{m}}$36.3$^{\rm{s}}$,
DEC$=$$-$42$^{\circ}$53$'$53.0$"$ (b$=$11$^\circ$
l$=$331.5$^\circ$). The positions of the centers of each individual
WFI CCD are seen in Table. 1. This particular field was chosen as
it has excellent visibility during the course of the run, it is
permanently located away from the moon, it lacks any stars brighter
than V$\sim$11 and is a highly crowded field with low dust
extinction. 
 
We require a signal-to-noise ratio of $\sim$200 at V$=$18.0 to adequately
recover the transits of Hot Jupiter planets. In order to do this, we
used a single broad-band filter covering the combined wavelength
range of Cousins V and R to increase the signal-to-noise. We took 5
minute exposures of the field, which with the V+R filter permitted a photon-noise signal-to-noise
of $\sim$220 for a 7-day moon for a V$=$18.5 star in 2$''$ seeing.
With a full moon this signal-to-noise decreases to $\sim$165.  A
total of 2201 images of the field were obtained over the course of 53
nights.  Given the CCD readout time, pointing and focusing overheads,
a mean temporal resolution of 6-7 minutes was achieved within our
observing window (typically 9 hours per night in good conditions). 
This combination of telescope, detector, filter and observational
strategy has been previously used successfully to perform a similar search for planetary transits
and variable stars in the globular clusters 47 Tucanae and $\omega$
Centauri \citep{W2004,W2005,W2006,W2007a}.

In addition to the V+R imagery, three images in V and three in I were taken with the same telescope and pointing in order to produce the field color magnitude diagram (CMD). Observations in V and I of Mark-A standard stars \citep{Lan1992,Stet2000} were also taken on the same night for accurate CMD calibration.

Image reduction was carried out using the standard MSCRED routines
within IRAF\footnote{IRAF is distributed by National Optical
Observatories, which is operated by the Association of Universities
for Research in Astronomy, Inc., under cooperative agreement with the
National Science Foundation.}. This included region trimming, overscan
correction, bias correction, flat-field correction and dark current
subtraction. Low quality images (poor seeing, poor focus, etc)  were filtered out, and the
resulting dataset used to produce precise, high resolution, photometric
time-series for as many stars as possible in the survey field.

\section{Lightcurve Production}
A total of 110,372 stars were identified in our survey field, and for
each we produced a photometric time-series using an application of Differential Imaging
Analysis (DIA), previously described as the optimal PSF-matching
package of \citet{AL98}. This code was subsequently modified by
\citet{Woz2000} for the detection of microlensing events. We direct
the reader to the Wozniak paper for a full description of the
code and its application.

By matching the stellar PSF throughout a large image database, the
systematic effects resulting from varying atmospheric conditions on
the output photometric precision is dramatically reduced. This method
allows ground-based observations the best prospects of detecting
small-amplitude brightness variations in faint targets. DIA is also
one of the optimal photometric methods when sampling fields with a
high degree of crowding, as the large number of stars permits a large
number of pixels to contain information on any PSF differences,
improving the PSF-matching process. Initial stellar flux measurements
are made via profile photometry on a template frame, produced via
median-combining a number of the best-quality images with small
offsets. The flux measurement on this template is used as the zero
point of the resulting stellar time series.

Stellar positions were found on a reference image, which contained the
best seeing conditions, and all the
subsequent data-set images, including the template, were
registered. The best PSF-matching kernel was determined, and each
registered image was subsequently subtracted from the template, with
the residuals generally being dominated by photon noise. Any object in
the frame that changed in brightness between the image and the
template was recorded as a bright or dark spot in the residual map.

Differential photometry does not automatically produce time-series in
magnitude units, rather in differential counts. This is a linear flux
unit output in the code from which a constant reference flux (taken
from the template image) has been subtracted. In order to convert to a
standard magnitude system, the total number of counts for each star
was measured using the PSF package of DAOPHOT within IRAF, with the
same images and parameters as used in the DIA code. The total database
of output time series were then converted in a standard way into
magnitude units via the relation:

\begin{center}
$\Delta m_i = -2.5 \log [(N_i+ N_{\rm ref,\it{i}}) / N_{\rm ref,\it{i}}]$
\end{center}

where $N_{\rm ref,\it{i}}$ is the total flux of star $i$ on the
template image and $N_i$ is the original difference flux in the time
series as produced with the photometric code.

The pixel coordinates of all the visible stars were determined
separately from the reference frame via DAOFIND within IRAF, with the
resulting profile photometry extracted from the subtracted frames at
those determined positions. Due to the filter chosen, the time-series are presented
here in V$+$R differential magnitude units. These can be converted to
the standard V system if required via the additional calculation of
color terms.
 
When combining differential fluxes with DAOPHOT-derived photometry on
the reference image, we must correct for errors based on the
individual apertures used. The scaling between the two fluxes was
determined via an aperture correction, which was performed on the
DAOPHOT magnitudes for the stars. This method is described in Appendix
B of \citet{H2004}. We found that our PSF magnitudes were consistently
0.06 magnitudes brighter than the aperture-derived values (using the
same aperture values as in the differential photometry). We corrected
for this by shifting our magnitude zero-point to
25.0$-$0.06$=$24.94. This correction ensures that the amplitude of
moderate variation detected by the DIA code is accurately represented in
magnitude units.

\subsection{Systematics Removal and Photometric Accuracy}
All 110,372 time-series as output by DIA were converted to $\Delta$V$+$R magnitude units as above. The whole dataset was cleaned for
systematic effects by running an application of the SYSREM systematics
removal code \citep{T2005}. This algorithm searches for and removes
effects common to the stars of a particular dataset, using only the
time of observation and the respective flux measurements for all stars
simultaneously. The result is a dramatic increase in the photometric
precision, of vital importance to any search for small amplitude
brightness variations. The algorithm works best for brighter stars,
where systematic effects dominate in the photometric uncertainties.

Fig.\space\ref{photplot} presents the output DIA-derived photometric
precision for 90,959 stars across the whole of the WFI field, after an
application of the SYSREM algorithm. These stars are those for which both a post-SYSREM lightcurve, V magnitude and astrometry have been derived. The V magnitudes for these stars were determined from the field color-magnitude diagram (see next subsection). The Y-axis shows the logarithm of
the measured root-mean-square ($\it{rms}$) uncertainty for the whole stellar
lightcurve, plotted against the V magnitude of the same star. Some
variable stars are visible as a separate sequence above the main $\it{rms}$
distribution (particularly for brighter stars). Also overplotted on
the figure is the theoretical photon noise due to the star (green
short-dashed line) and the sky noise contribution (blue long-dashed
line). A further noise contribution arises from residual image-based systematic
effects; flat-fielding errors, scintillation and CCD non-linearity
response. All of these are estimated with an amplitude of four times
the calculated sky noise to represent the total observed noise when
added to the photon noise in the star and the sky, and is displayed on
Fig.\space\ref{photplot} as a dash-dotted blue line. The solid red
line represents the total of the photon noise and this total sky
contribution (added quadratically), and describes the photometric
uncertainties well.  Our photometry is photon-noise dominated to a V magnitude of $\sim$18.5, where it becomes sky+residuals dominated to fainter magnitudes.

\subsection{Color-Magnitude Diagram and Astrometry}
A color-magnitude diagram (CMD) of V$-$I against V was produced for the survey
field with the same telescope and detector, in order to place the detected variables (and transit candidates)
onto the standard magnitude system. The diagram contains a total of
95,358 stars with both measured V and I magnitudes. This CMD, with all 494
detected variables overplotted, can be seen in
Fig.\space\ref{cmdplot}. The figure shows four panels, one for each of
the main type of variable star found (overplotted). The four main
types can be seen to populate different regions of the CMD, depending
on their intrinsic spectral type. Our survey has a saturation limit at
V$=$15.0, and a faint limit of V$=$22.0. A sharp edge is seen at
V$-$I$=$0.6, indicating the Galactic disk main sequence turn-off, the limit of the more numerous F,G,K and M stars (to the right of the edge) compared to the far less numerous
earlier type stars (to the left of the edge) at the typical age of the galactic disk. The DAOPHOT-derived errors and calibration uncertainty in our magnitude determinations are
also overplotted.

The CMD was calibrated via V and I observation of the MarkA standard stars
\citep{Lan1992,Stet2000}, taken at the same time as the CMD data. A total of 415 standards were
cross-identified in the field via matching of astrometry\footnote{MarkA
astrometry and photometry downloaded from
http://www4.cadc-ccda.hia-iha.nrc-cnrc.gc.ca/community/STETSON/standards/}. The
mean and standard deviation of the shift in magnitude was found for
each CCD and for each filter independently. The resulting calibration
uncertainty is 0.03 magnitudes in V and 0.05 in V$-$I. As an
additional check, the CMD was also cross-correlated with 2,000 stars
within 20$'$ of the field center in the NOMAD online catalog
\citep{Zach2004}, and, being towards the faint limit of NOMAD, have V
magnitudes within 0.2 magnitudes (1$\sigma$) of the NOMAD results.

The astrometry for all of the stars in our database was determined via a
search of the USNO CCD Astrograph Catalog (UCAC1), to search for astrometric standard stars within the field. Several
hundred such stars were successfully identified, producing an accurate
determination of the astrometric solution for the stars in each CCD
independently; the resulting calibration accuracy was 0.25$''$ (0.66 pixels). The
extent of our field can be seen in Fig.\space\ref{astromplot}, plotted
as $\Delta$RA and $\Delta$Dec in degrees. The
location of the variable stars are overplotted with the same color
scheme as in Fig.\space\ref{cmdplot}. The non-functioning WFI CCD7 is
clear.

\section{Variability Search Method and Completeness}
We used the `analysis of variance' (AoV) statistic to perform the
variability search, with \citet{S1989} providing a full description of the
method. Via AoV, the data are phase-wrapped to a trial period and
grouped into phase bins. A one-way statistical analysis of variance is
performed on the result with the output noted. This procedure is
repeated for a fixed range of test periods for each star, producing a
series of significances and their corresponding periodicities. The
final output for each star is the peak periodicity and its
corresponding significance.

The output AoV significance is far higher if a periodicity is
detected, facilitating the variable identification, and allowing for
detection thresholds to be used to increase the search
efficiency. Fig.\space\ref{aov_hist} shows the output AoV significance
(S$_{\rm{AoV}}$) distribution for the whole time-series dataset. The main
population at low significances are those stars for which no
periodicity was detected. The variable star candidates constitute the
long tail to increasingly higher significance. The main population was
fit with a Gaussian, overplotted in Fig.\space\ref{aov_hist} with a
resulting mean S$_{\rm{AoV}}=$3.904 and standard deviation ($\it{rms}$) of
S$_{\rm{AoV}}=$0.499.

Any candidate with a significance $\ge$3$\times$$\it{rms}$ (9,230 candidates
S$_{\rm{AoV}}$$\ge$5.401) was counted as a candidate variable star, which
was then phase-wrapped at the AoV determined peak periodicity and the
presence of any variability noted. This candidate detection threshold
is marked in Fig.\space\ref{aov_hist} with a vertical line.

We visually searched through all 9,230 candidates to detect the
variable stars, which were also phase-wrapped to 0.5, 2 and 3 times
the detected period to check for the presence of integer
aliasing. Out of these, we visually detected a total of 572 variables, corresponding to one real
variable per 16.1 candidates. With such a large number, the chance of any real
variables being missed in the search and escaping detection via this
method is low.

The total candidate list numbered 572 stars that were seen to undergo periodic brightness changes, 77 more than the final variable list. The
astrometry for the original 572 were checked for double entries,
consisting of candidates that lie within several arcseconds of
each other in both RA and DEC, and also variables with the same or very
similar ($\le$0.1d) peak periodicity. The majority of the extra 77
were found to be double entries, particularly comprised of fainter
`variables' which lie close ($\le$5$''$) to a far brighter real
variable. We classify these as `blends', and have lightcurves with the
same periodicity as the bright nearby variable, but with only a
fraction of the signal. The RR Lyrae stars, being so bright, were particularly
responsible for these blended candidates. The remaining few false
variables were identified by their common periods to more distant, yet
still nearby, bright counterparts. In all, 494 true variables remained
in the candidate list and constitute the final catalog.

A small number of the detected variables ($\sim$5$\%$) had significant
scatter on the resulting phase-wraps, indicating an incorrect
determination of the period. These were then analyzed with a
Lomb-Scargle Periodogram \citep{Brett01} to provide a secondary
estimate to the period. This was then phase-wrapped and the period optimized
until a minimum in the scatter was obtained. The change in
period needed to obtain the final period was less than 0.1d in all
cases, and is related to the number of datapoints contained within the periodicity. The small number of these statistical false-period detections
is further evidence of the strength of AoV as an accurate periodicity
determinator.

Fig.\space\ref{amplitudes} presents the total amplitude of the
variation for each cataloged variable star, plotted as a function of
its corresponding V magnitude. This provides information on the
detection limits of the catalog. The dotted line defines the observed
empirical limit in the variable amplitude as magnitude increases. Any variable
with an amplitude less than or equal to the position of the dotted
line is unlikely to be revealed by the detection method described above. Hence,
at V$=$18.0, the observed detection limit is 0.05 magnitudes
(4.7$\%$), while at V$=$22.0, the limit is 0.70 magnitudes (91$\%$),
and originates from our photometric uncertainties (Fig.\space\ref{photplot}).

\section{Variable Star Catalog}
In total the catalog contains 494 variable stars. This consists of 190
eclipsing binary (EcB) systems, displaying many examples of detached
(eg: V8), semi-detached (eg: V26) and contact (eg: V3,)
configurations. It also contains 51 pulsating stars (puls) defined as
having a sinusoidal variation and period $\le$2d; and 237 long period
variables (LPV), which display similar variation to the pulsators but
with far longer periods ($2 \le P \lesssim 100$d). We also detected 11
$\delta$ Scuti stars with extremely short pulsation periods (minutes
in many cases) and 4 field RR Lyrae stars (three type `AB' and one
type `C'). We also detected a single irregular variable. The
lightcurve database is available on the online edition of this paper.

The total variable star catalog (in order of RA) can be seen in
Table\space2. This table presents the identification number of the
variable, the type of variability (tabulated in order of type: EcB, RR
Lyrae, $\delta$ Scuti, puls, LPV and Irr), the period in days, the RA
and DEC (J2000.0), the V magnitude, V uncertainty and finally the V$-$I color
and associated error. The V magnitude, uncertainty and V$-$I color are all taken from the color magnitude diagram dataset. The uncertainties include calibration errors. If a particular binary has an orbital period
$\le$0.385d (from the determined 3$\sigma$ lower limit to their period
distribution, see later) we classify it as a W UMa type system, and is
marked as such in the table. Those magnitudes marked with `$-$' have unknown magnitude in that particular band, the vast majority of which are due to saturation in one filter.

Accurately determined periods are presented to five decimal
places. Variables with less certain periods are presented with a
number of decimal places appropriate to the uncertainty. Those
pulsating variables which could in reality be short-period eclipsing
binaries, indistinguishable in our data, with equal-mass components W Ursae Majoris (W UMa) stars, and hence identical
primary and secondary eclipses, have their periods marked with an
asterisk. Those longer period variables and binaries with only one
single eclipse visible (hence uncertain periods) have their estimated
periods presented, which are marked with a `$\sim$'. Those
variables classed as pulsators which could be actually eclipsing
binaries with twice the tabulated period are marked with a `$\ast$'.

The total catalog of variables is overplotted on the dataset CMD in
the four panels of Fig.\space\ref{cmdplot}, to facilitate
determination of their likely nature. The EcB stars (top left, blue squares) are
seen to inhabit all parts of the diagram, indicating the differing
component spectral types and distances to these systems. In contrast
the $\delta$ Scuti stars (bottom left, yellow circles) and RR Lyrae (bottom left, green
triangles) are all located blueward of the main `shoulder' seen at
V$-$I$=$0.6. This indicates that these types of variability are
associated exclusively with stars of earlier spectral types
($\leqslant$A-F). No faint $\delta$ Scuti stars were seen in our data,
due to the low amplitude of their variability. In comparison, all but two of the LPV (V200 and V401)
(red triangles) are located redward of the V$-$I$=$0.6
shoulder, indicating the association of this type of variability with
later spectral types ($\geqslant$G$-$M). All but four of the pulsators (bottom right; V291, V319, V338 and V491)
are located red-ward of V$-$I$=$0.5.

\subsection{Eclipsing Binaries}
A total of 190 EcB systems were found in our data. These include many
examples of contact, semi-detached and detached systems. The vast
majority of the binaries have accurate periodicities determined, with
the exception being those long period fully detached systems with only
one or two eclipses visible during our observing run. We classify those 17 binaries blue-ward of V$-$I$=$0.5 (V36, V44, V76, V113, V117, V168, V228, V253, V257, V294, V296, V317, V385, V405, V442, V467 and V476 as seen in Fig.\space\ref{cmdplot}) as candidate blue-stragglers. Particularly, V385, V405 and V467 have `blue' V$-$I magnitudes of 0.244, 0.272 and 0.253 respectively. Some binaries (for example V10 and V31) are difficult to phase-wrap accurately, due to the low number of eclipses visible in our data.

\subsubsection{Selected Binary Systems}
Several of our binary systems are of particular interest for follow-up
studies. V413 is the brightest binary detected in this search, with
V$=$14.595$\pm$0.008 and orbital period 1.7962d. The total
phase-wrapped lightcurve for this semi-detached system is seen within
Fig.\space\ref{varplot7} and in greater detail in
Fig.\space\ref{V413}. It displays a very deep primary eclipse
($\Delta$V$+$R$\sim$2 mags) and a far shallower secondary eclipse. This indicates that the two stellar components are of very
different spectral types and luminosities.

Closer examination of the V413 lightcurve reveals a
secondary pulsation variation with a period of only a few minutes, typical of a $\delta$ Scuti star. The amplitude of this pulsation is $\geqslant$5 mmag and can be seen in
both panels of Fig.\space\ref{V413} (phase-wrapped in the bottom panel to 5 times the periodicity for ease of visibility). Such variations indicate the presence of a small-scale surface variability similar to solar surface
oscillations. The period of this secondary oscillation is comparable
to the exposure time of the observations, hence the amplitude
presented here carries uncertainty. Furthermore, the period of this
oscillation is seemingly 1/500$^{\rm{th}}$ the binary orbital period,
perhaps indicating that it may originate with surface jitter induced
by the binary companion. A second theory, and perhaps more likely, is that one component of this binary is a $\delta$ Scuti star. Follow-up photometry and spectroscopy are planned to
fully understand this interesting system.

V161 displays a long period sinusoidal variation with a superimposed
eclipse with the same period. The system is quite red (V$-$I$=$0.96)
and is perhaps composed of a pulsating red variable, which is in turn orbited by a
stellar companion. Alternatively the system could contain a red giant which has
been significantly distorted by a nearby massive companion.

As our survey field was observed for 27 nights in June 2005 and 26
nights in June 2006, changes in any variability during the course of
that year can be seen in our lightcurves. V130 (orbital period
1.22086d) displays a consistent change in system brightness
(out-of-eclipse zero-point) of $\sim$0.1 mags (V+R) during this
time, the system is brighter in 2007. This can be seen in the phase-wrapped lightcurve for this system
in Fig.\space\ref{varplot2}. We attribute this variation to one
component being a long period variable in itself. As the change is stable over the course of a month, we cannot attribute this to starspots or flare activity. If the components are
interacting with each other, as the out-of-eclipse ellipsoidal
variations seem to indicate, then an accretion scenario may be present
in this system.

A number of our binaries are likely composed of low-mass M-dwarf
components, as such they are of importance to studies of stellar mass
and radius relationships for late-type stars. These systems were
chosen based on their V magnitude and V$-$I color, along with their
short orbital periods and the evidence for only a small degree of
interaction between the components in their lightcurves. We classify
V26, V115, V184, V297, V304, V352, V355, V402, V418 and V438 as likely
low-mass systems. Particularly interesting are V304 and V402 as
detached systems. V232 is extremely faint (V$=$21.99) with a V$-$I of
2.91, and is almost certainly composed of very low-mass stars.

A number of the longer period systems could also contain a low-mass
component again based on the amplitude of gravity effects and the
difference in primary and secondary eclipse depths (for example V144 and
V312). Indeed, the systems classified as W UMa stars (seen in
Table.\space2) almost certainly contain low-mass
components. Interestingly these W UMa stars are seen as mostly blue straggler
systems in old globular clusters \citep{W2004,W2007a} rather than
late-type stars expected for the (younger) general field. The shortest
period binary detected in this survey is V344, with an orbital period
of 0.2009 days (V$=$20.9), and if confirmed will become the
shortest period binary known (beating shorter than the previous 0.215d binary found
in 47 Tuc by \citealt{W2004}).

\subsubsection{Eclipsing Binary Period Distribution}
Fig.\space\ref{binaries} displays the {\it{n(P) dP}} eclipsing binary period distribution, to an maximum limit of 1d, which contains all the binaries we define as contact binaries. The number of binaries can be seen to generally increase as period decreases, with a population of binaries centered at a period of $\sim$0.27d. We define these as the W Uma systems, short period equal-mass contact binaries that make up the most commonly observed type of field binary (L.Kiss, private communication).

Given the very short periods of these systems ($\sim$0.27d), the total length of our observing window (53 nights or 100 times the typical orbital period assuming a night is 0.5d long), and the considerable amplitude of these
systems ($\ge$0.1 magnitudes), we conclude that all systems with suitable inclination to undergo eclipses will have been detected in our field down to a magnitude of V$=$18.8, as determined from our photometric uncertainties (Fig.\space\ref{photplot}) and our total catalog completeness limits (Fig.\space\ref{amplitudes}). They do not suffer in any significant way from detection-related bias. We take the Gaussian appearance of our W Uma period distribution as further evidence of its completeness.

We find a best-fitting Gaussian (via least-squares residuals) to our W UMa population (overplotted on Fig.\space\ref{binaries}) to have a mean orbital period of 0.277d with associated standard deviation 0.036d. This result compares very well to the 0.27d mean volume-limited W UMa period as determined by \citet{Ruc2007}, from field binaries in the All Sky Automated Survey. We hence determine the 3$\sigma$ lower limit to our W UMa periods to be 0.17d. Our shortest period binary is 0.2009 days, and is 2.1$\sigma$ below our mean. W UMa binaries with a period equal to or less than this are exceedingly rare.

\subsubsection{Binary Frequencies and comparison to Globular Clusters}
In the haloes of the globular clusters 47 Tuc and $\omega$ Cen, \citet{W2004,W2007a} determined an observed eclipsing binary frequency of 1.7$\pm$0.4$\times$10$^{-4}$ and 5.3$\times$10$^{-4}$ respectively (the cluster cores are saturated). The value for 47 Tuc is substantially lower than that observed in the cluster core \citep{A2001}, and this has been taken as evidence for mass segregation in 47 Tuc. Binary destruction due to cluster dynamical processes would produce a substantially higher binary frequency in the general Galactic field.

In our Lupus field, we determine an eclipsing binary frequency of
190/110372$=$1.7$\pm$0.4$\times$10$^{-3}$, with the error being
3$\sigma$, derived from the Poisson error in both the binary and total
star number and applied with standard quotient error propagation. This
equates to one detected binary per $\sim$600 stars, ten times the
value for 47 Tuc and three times the value for $\omega$ Cen.

This strengthens the argument that binaries are less common
(particularly long period detached binaries) in the outskirts of globular clusters, being mainly due to mass segregation, and also due
to binary destruction. Mass segregation is not thought to be
prevalent in $\omega$ Cen (which has three times the observed binary frequency of 47 Tuc, yet only 1/3$^{rd}$ that of our Lupus field). All three
surveys contain a similar total number of stars (within 10$\%$) and
have very similar detection thresholds, due to similar global dataset
properties. Clearly, the general galactic field contains a higher rate
of detectable binary stars than is contained within the haloes of
these two globular clusters.

\subsection{$\delta$ Scuti and RR Lyrae stars}
A total of 11 $\delta$ Scuti stars and four RR Lyrae stars (three type
`AB' and one type `C') were detected in our data. The $\delta$ Scuti
stars (also known as a Dwarf Cepheids, AI Velae stars or AI Velorum
stars) undergo both radial and non-radial pulsations on short
timescales. They were classified as such in this work from their
pulsating nature, bluer V$-$I values and very short periods. The
tabulated data for these variables can be seen in Table.\space2 and
their lightcurves are presented in Fig.\space\ref{varplot8}.

The determined periods range from 0.003919d (V459, 5.6 minutes) to
0.08953d (V270, 2.1 hrs). They (along with the RR Lyrae) are located
blueward of the main stellar type `shoulder' seen in
Fig.\space\ref{cmdplot}, indicating their classification as stars of
earlier spectral type (typically A to F). The moderate amplitude of
the V459 pulsation precludes it from being composed of solar-like
convection-driven surface oscillations. From the total stellar
database of 110,372 stars, we estimate an apparent $\delta$ Scuti
occurrence frequency in our field of 11/110,372$=$10.0$\pm$0.3$\times$10$^{-5}$. This corresponds to one $\delta$ Scuti per 212 arcmin$^{2}$.

RR Lyrae are pulsating horizontal branch stars, that are generally old
and with relatively lower mass ($\sim$0.8M$_{\odot}$). Their
identification in the field (particularly in large field surveys
targeting the Galactic halo) is important to trace the extent of halo
streams. Three of the four detected here are very likely disk stars, due to
their apparent brightness, and have pulsation periods and lightcurve
shapes typical of the two main classes of RR Lyrae
\citep{Vivas01}. Similarly their data are presented in Table.\space2
and Fig.\space\ref{varplot6}. None show evidence of the little
understood Blahzko Effect \citep{Bla07}. We estimate an apparent
occurrence frequency of 4/110,372$=$3.6$\pm$0.1$\times$10$^{-5}$ ($\sim$1 RR Lyrae
per 583 arcmin$^{2}$) for field RR Lyrae towards Lupus. None of the
rare AHB RR Lyrae \citep{Sand94} were seen in our data.

RR Lyrae are a well known class of variable due to their usefulness as
distance indicators, however their absolute magnitude depends on their
metallicity \citep{Sand1981a,Sand1981b}. From studies of RR Lyrae and
horizontal branch stars in the Milky Way, the LMC and M31 clusters
this relation has been adopted with a slope 0.20-0.23 mag/dex
\citep{Cl2003,Ca2003,G2004}. \citet{R2005} adopted a relation of the
form $M\rm{_{V}}$$=$0.20$\pm$0.09[Fe/H]$+$0.81$\pm$0.13, which was
also used to estimate the distance to $\omega$ Cen by
\citet{W2007a}. The metallicity must be determined spectroscopically
before the distance can be estimated with any certainty.

However, if we assume solar metallicity ([Fe/H]$=$0) for all four RR
Lyrae, using the relationship above we determine an absolute V
magnitude of 0.81$\pm$0.22. By applying the maximum degree of
reddening for the field (E(B$-$V)$=$0.182 mag \citep{Sch1998}), the V
magnitudes of the stars would be altered by
V$_{\rm{obs}}$$-$3.315$\ast$E(B$-$V). Hence for the four RR Lyrae and their
corresponding V magnitudes (V$_{\rm{obs}}$) as listed in
Table.\space2, we derive minimum reddening-corrected apparent V
magnitudes of 14.602$\pm$0.003, 15.063$\pm$0.002, 14.907$\pm$0.002 and
17.146$\pm$0.005 for the four stars.

We determine distance modulus (m$-$M)$_{0}$ upper limits of
13.80$\pm$0.22, 14.25$\pm$0.22, 14.10$\pm$0.22 and 16.34$\pm$0.22
($\sim$5.7, $\sim$7.0, $\sim$6.6 and $\sim$18.5 Kpc) for our four RR
Lyrae, with equal uncertainties throughout as they are strongly
dominated by the uncertainty in the absolute magnitude. The faintest
RR Lyrae (V387) is likely a background Halo star. Actual
metallicity measurements will permit more accurate distances for these
stars, rather than the upper limits presented here.

\subsection{Miscellaneous Pulsators}
Our catalog also contains 51 pulsating stars of various types and one
irregular variable. We define the pulsators are stars displaying
regular pulsation brightness variations with periods less than two
days. The global properties of these stars are presented in
Table.\space2, and the lightcurves for the first 20 pulsators (for
illustrative purposes, as visually they appear very similar) can be seen in
Figs.\space\ref{varplot9}-\ref{varplot10}. Longer period pulsators are
classified here as long period variables (see below). Their position
on the CMD is mostly scattered (Fig.\space\ref{cmdplot}) although the
majority are quite red, indicating their makeup as later spectral
types. The irregular variable (V134) is quite blue (V$-$I$=$0.35)
indicating it likely is of A-F spectral type and may contain both
radial and non-radial pulsations. Their apparent occurrence frequency is 4.6$\pm$0.3$\times$10$^{-4}$ (one pulsator for every 46 arcmin$^2$).

The majority of the pulsators display regular radial sinusoidal
variations, with periods in the range 0.4$\rightarrow$0.6 d, with the
lowest being 0.12377d (V238). We classify them as `pulsators', or red variables, being stars
passing through the instability strip during their later
evolution. V295 has a period of only 0.04159d, a typical period for a
$\delta$ Scuti star, however the V$-$I color is 0.932, unusually red
for a $\delta$ Scuti, hence its inclusion here.

\subsection{Long Period Variables}
We also detected a total of 237 long period variables (LPVs), by far
the most common type of variable seen in our Lupus field, with an observed occurrance frequency of 2.2$\pm$0.3$\times$10$^{-4}$ (one LPV per 10 arcmin$^2$). We define
these as displaying regular sinusoidal (or slightly saw-tooth)
variations with various periodicities greater than an arbitrarily chosen lower
limit of two days. The global properties of these stars are also seen towards the end
of Table\space2. The LPVs are in the vast majority very red stars,
indicating their likely nature as red giant stars of late spectral
types. The lightcurve data themselves can be accessed as online data.

Many of the periods presented in Table.\space2 are uncertain, and
marked appropriately with a `$\sim$'. This is due to the limitations
imposed by our 53-night dataset, variables with significantly longer
periods (Mira variables) would be flagged as variable but
have indeterminate final periods. Further photometry spanning a long
baseline (months) is needed to determine accurate periods for these
stars, and is outside the scope of the this project. The lightcurve
data (in ASCII format) for all 494 variables in the catalog are
available on the on-line edition of the Journal.

\section{Conclusions}
We present a variable star catalog, along with a preliminary analysis of the stars, detected during a long-baseline (53 nights: 26 contiguous nights in June 2005 and 27 in June 2006) high temporal resolution (6-7 minutes) photometric survey to detect transiting short period planets towards Lupus. A total of 494 variable stars were detected in our 0.66 deg$^{2}$ field, all of which are new discoveries. The catalog comprises 190 eclipsing binaries (of contact, semi-detached and detached configurations), 51 miscellaneous pulsators, 237 long period variables, 11 $\delta$ Scuti stars, 4 field RR Lyrae and a single irregular variable. 

We determine a period distribution of 0.277$\pm$0.035d for our detected short period W UMa binaries (to V$=$18.8). Several binaries appear to be composed of equal-mass M-dwarf components, and others display evidence of mass transfer. We detected 17 candidate blue straggler stars and we also detected a binary with period 0.2009d (V=20.9), which constitutes the shortest period eclipsing binary known. Our brightest binary likely harbors a $\delta$ Scuti component.

We have used this catalog to determine the occurrence frequencies of the various types of variable in our field, and compared the overall frequency of detected binary stars to that previously determined for the haloes of the globular clusters 47 Tuc and $\omega$ Cen. We find that the
frequency of binary stars is significantly larger in our field than that found in either cluster (ten times for 47 Tuc and three times for $\omega$ Cen). This favors the scenario in which binaries are readily mass segregated into the cluster cores and/or destroyed by cluster dynamical processes.

\section*{Acknowledgments}
The Authors wish to thank the following people for their assistance during this work: Penny Sackett, Brandon Tingley, Grant Kennedy and Karen Lewis for their help with various aspects of this survey, and Laszlo Kiss for helpful discussion on W UMa stars. We also thank the anonymous referee for their detailed and helpful comments in improving this manuscript.


\begin{center}
\plotone{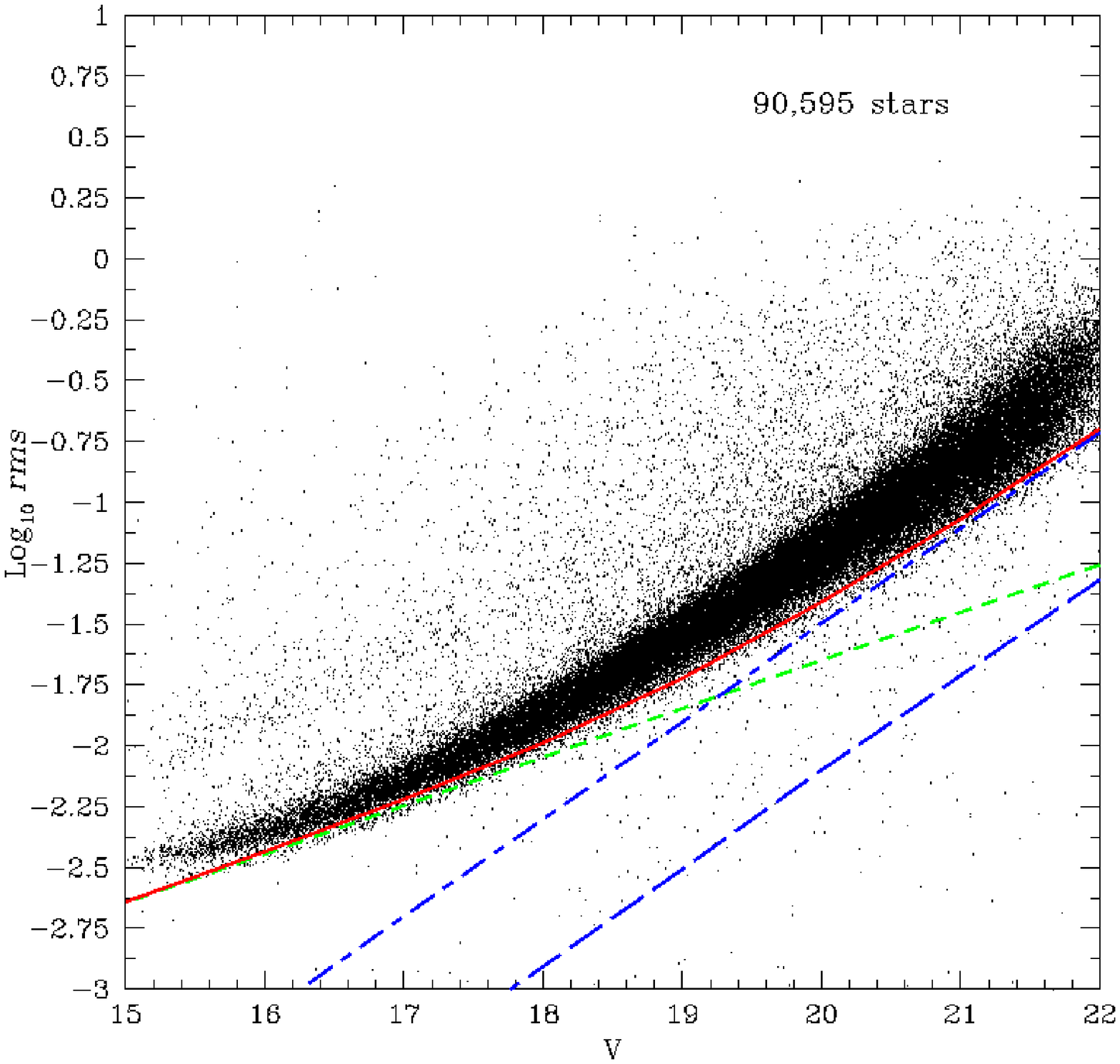} \figcaption[weldrake_fig1.eps]{The photometric precision of 90,959
stars across the field resulting from the DIA+SYSREM photometry technique applied to
our dataset, plotted against V magnitude. The theoretical photon noise
for the star (green short-dashed), the sky (blue long-dashed), the
residual noise contribution (blue dot-dashed) and the sum of all (red
solid line) are overplotted for comparison. Some variable stars are visible as the sequence of higher
$\it{rms}$ at bright magnitudes.\label{photplot}}
\end{center}

\begin{center}
\plotone{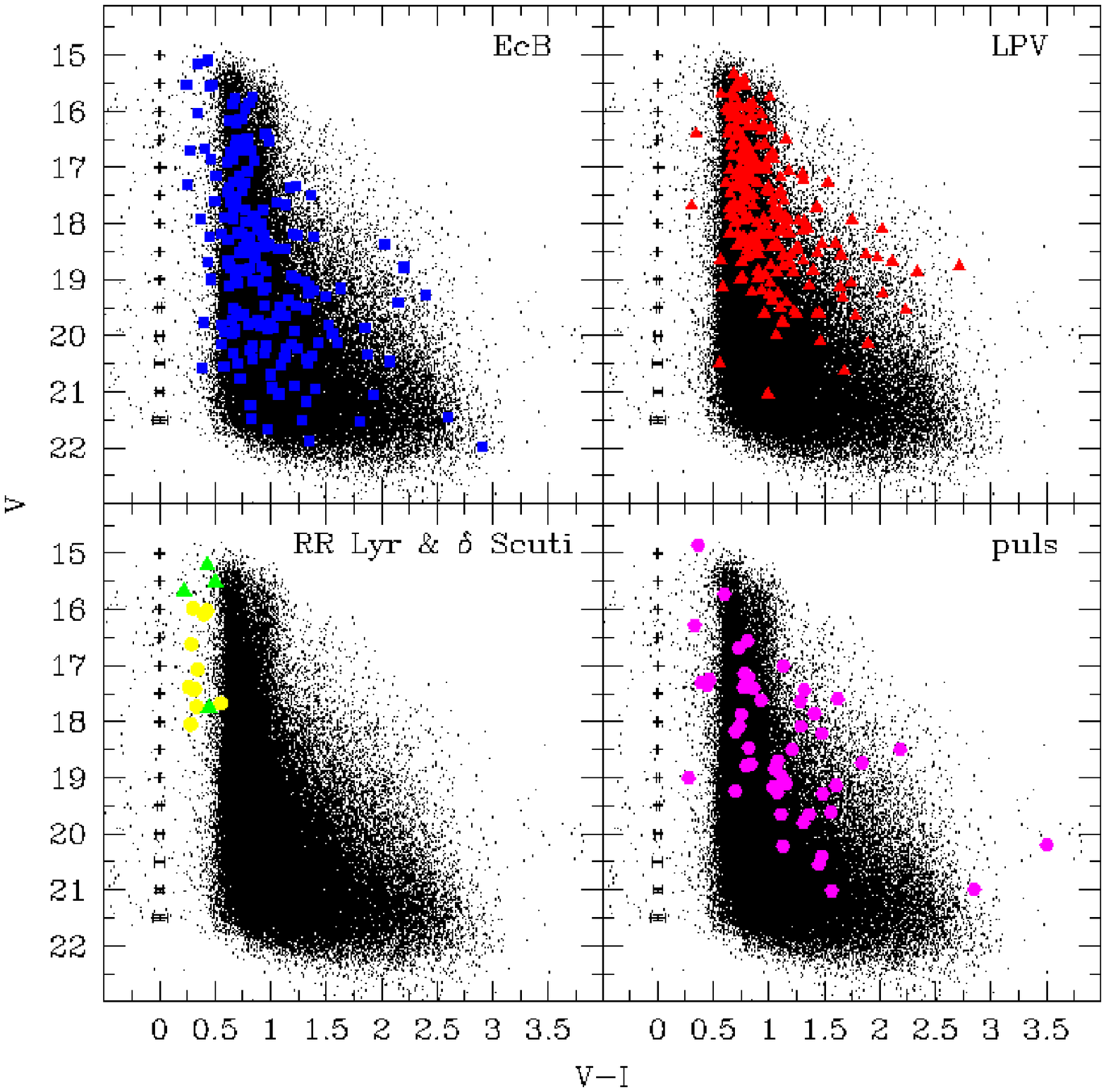} \figcaption[weldrake_fig2.eps]{The Color-Magnitude
Diagram (CMD) for our Lupus field. The locations of all 494
detected variable stars are overplotted. The four panels denote the
four main types of variable in the catalog, plotted in various colors
to facilitate their visibility. The various types of variable are seen
to populate various regions of the CMD. The $\delta$ Scuti (yellow) and RR
Lyrae stars (green) are all blue-ward of the main shoulder defining the limit
of A-F type Galactic Disk stars, indicating their early spectral types. The EcB are scattered all over the plot, indicating the different spectral types of the stellar
components. The 17 binaries with V$-$I values $\le$0.5 are candidate blue stragglers. The overplotted errorbars represent the output DAOPHOT
errors and the CMD calibration uncertainty added in
quadrature.\label{cmdplot}}
\end{center}

\begin{center}
\plotone{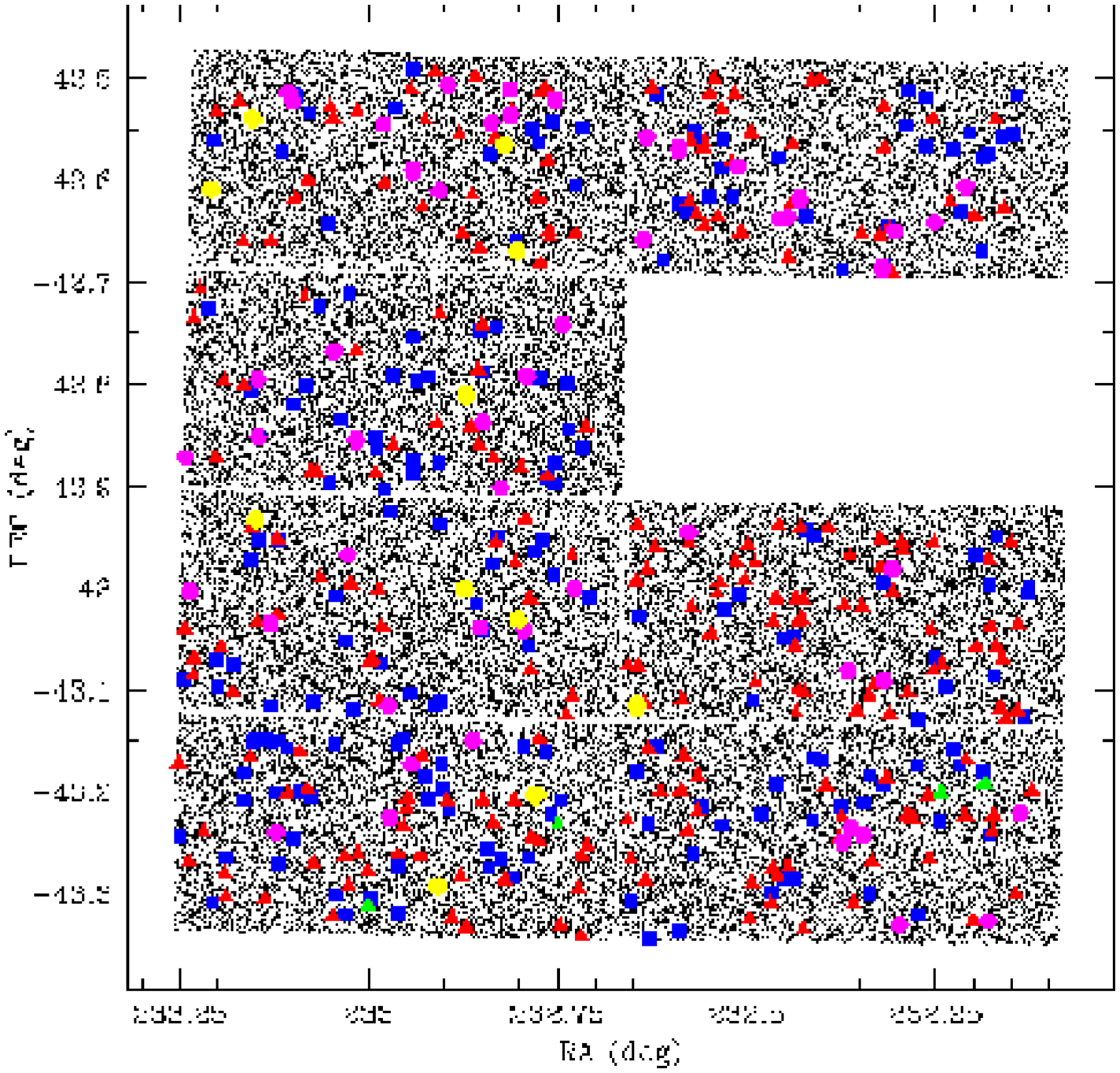} \figcaption[weldrake_fig3.eps]{The astrometry
for our field. The spatial distribution of the main types of
variable are overplotted (with the same color schemes as in
Fig.\space\ref{cmdplot}). The non-functional CCD7 is clear.\label{astromplot}}
\end{center}

\begin{center}
\plotone{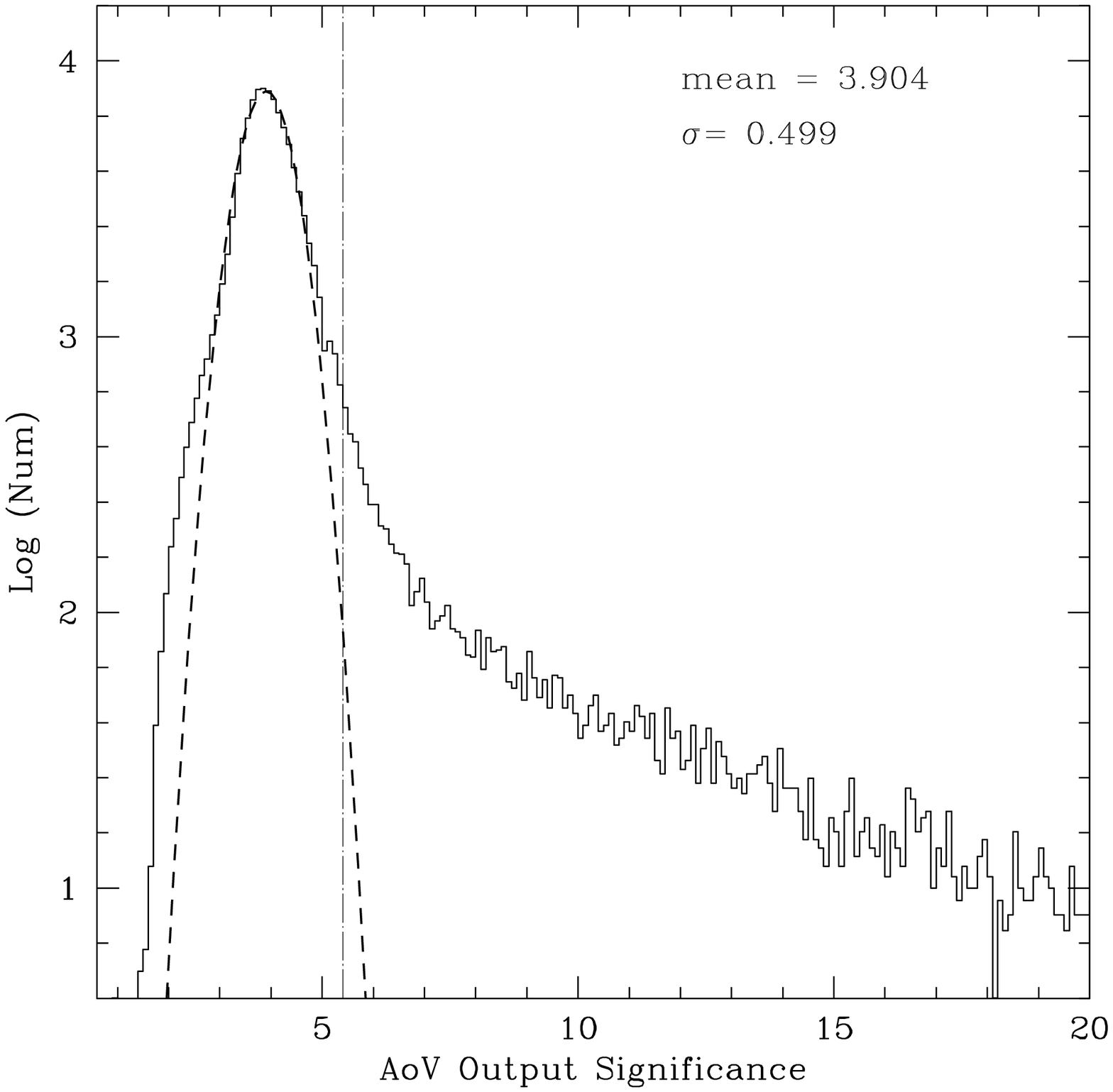} \figcaption[weldrake_fig4.eps]{The output AoV significance (S$_{\rm{AoV}}$) distribution for the whole 110,372 stellar lightcurve
database upon which it was run. The main population at lower
significance defines the general population for which no periodicity
was detected. A Gaussian was fitted to this population and overplotted
along with the resulting mean and standard deviation parameters. Any
lightcurve that had an AoV output significance $\ge$3$\times$ the
standard deviation of the background (marked as a vertical line) was
flagged as a candidate variable. All 9,230 candidates satisfying this
criteria were searched for variability.\label{aov_hist}}
\end{center}

\begin{center}
\plotone{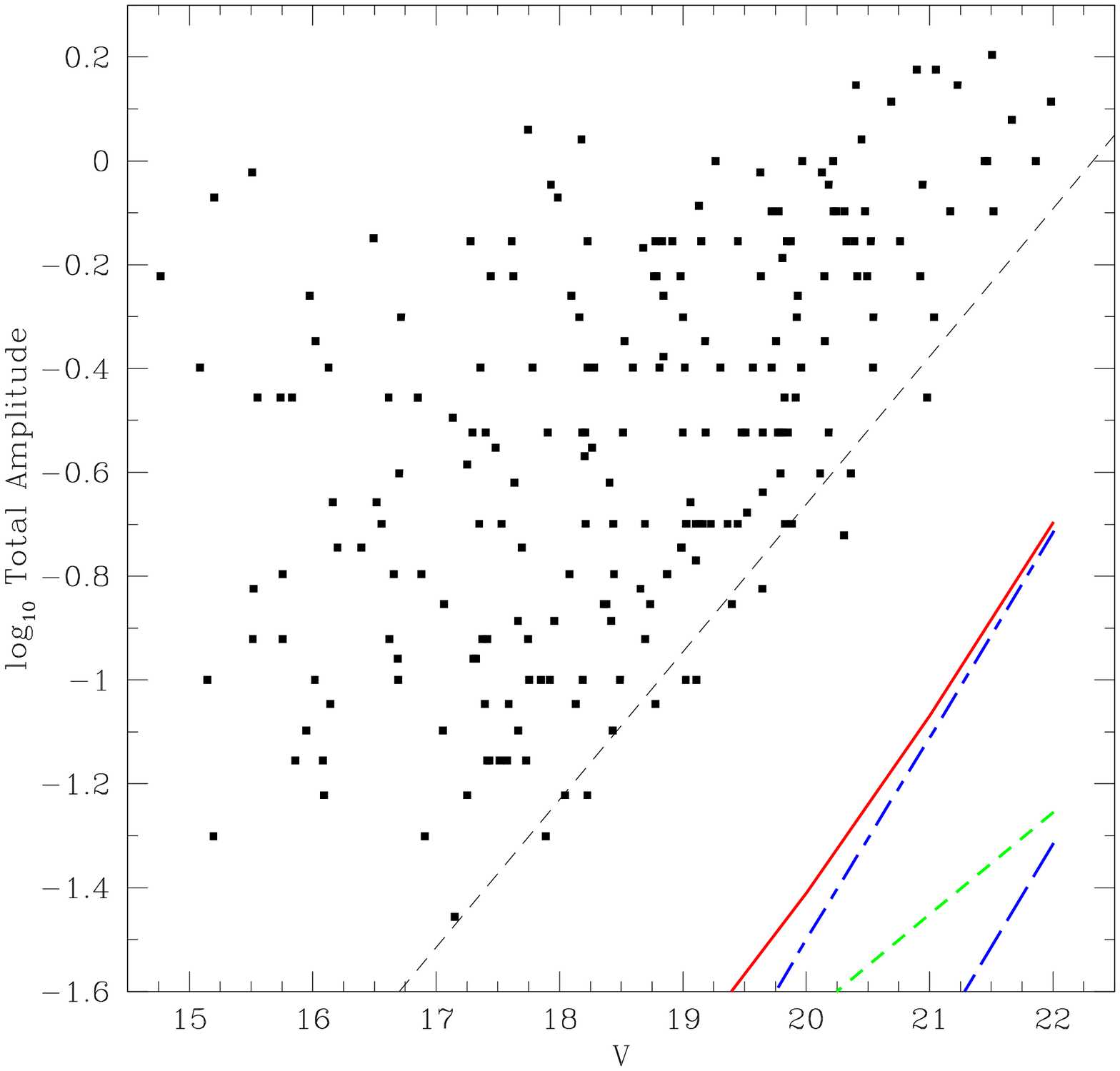} \figcaption[weldrake_fig5.eps]{The total V+R amplitude
for all variables presented in Figs.\space\ref{varplot1}-\ref{varplot10}, plotted against their V
magnitude. The empirical detection limit of our catalog is marked by the dotted
line. Any variables with an amplitude below this line are unlikely to be detected
in our survey. Also overplotted are the photometric noise contributions as per the color scheme of Fig.\space\ref{photplot}.\label{amplitudes}}
\end{center}

\begin{center}
\plotone{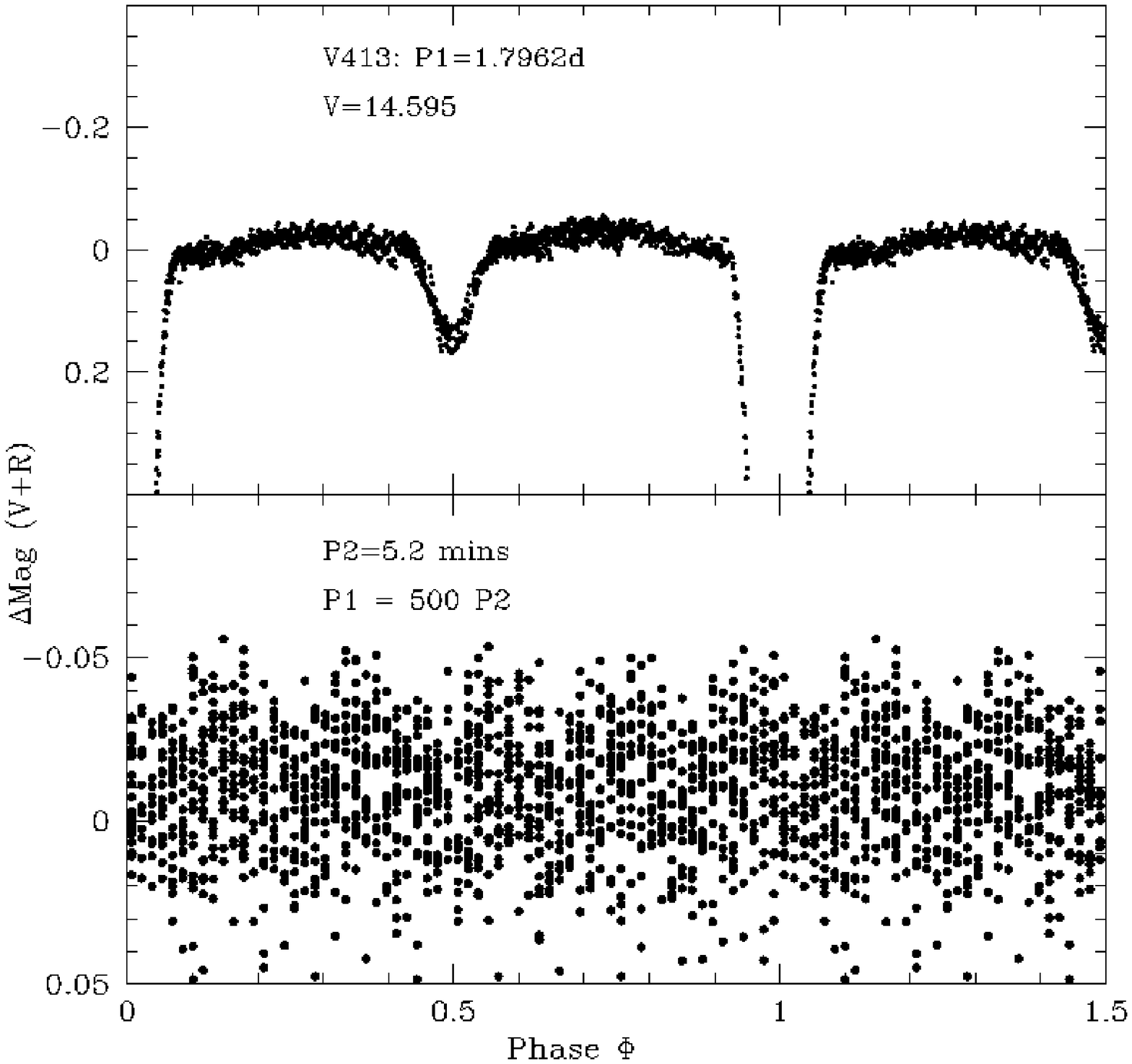} \figcaption[weldrake_fig6.eps]{The small-scale pulsation
variations of V413, as seen superimposed on the main binary lightcurve
(top), and in close-up (bottom), phased wrapped to five times the 5.2 minute
periodicity for ease of visibility. Such oscillations, present on the faint component of the
binary, are perhaps induced by the companion, or indicative of a $\delta$ Scuti component.\label{V413}}
\end{center}

\begin{center}
\plotone{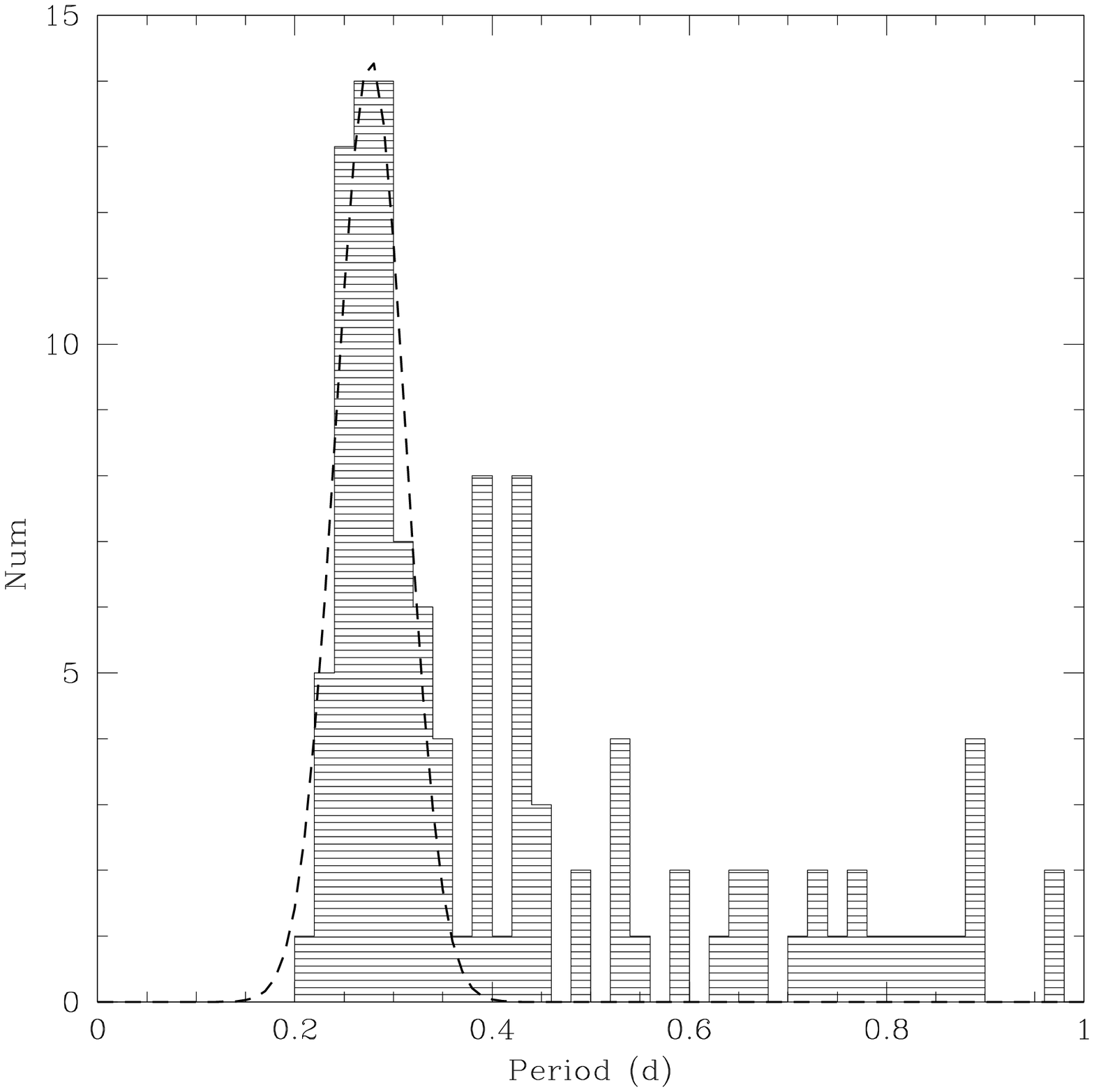} \figcaption[weldrake_fig7.eps]{The
{\it{n{P} dP}} period distribution of all detected eclipsing binaries with
period $\le$1d. The shortest periods are well represented by a
Gaussian with mean 0.277d and rms 0.036d. This population constitutes
the W UMa stars. A decreasing tail to longer periods is also seen,
which we attribute here to non W UMa contact and semi-detached
binaries.\label{binaries}}
\end{center}

\plotone{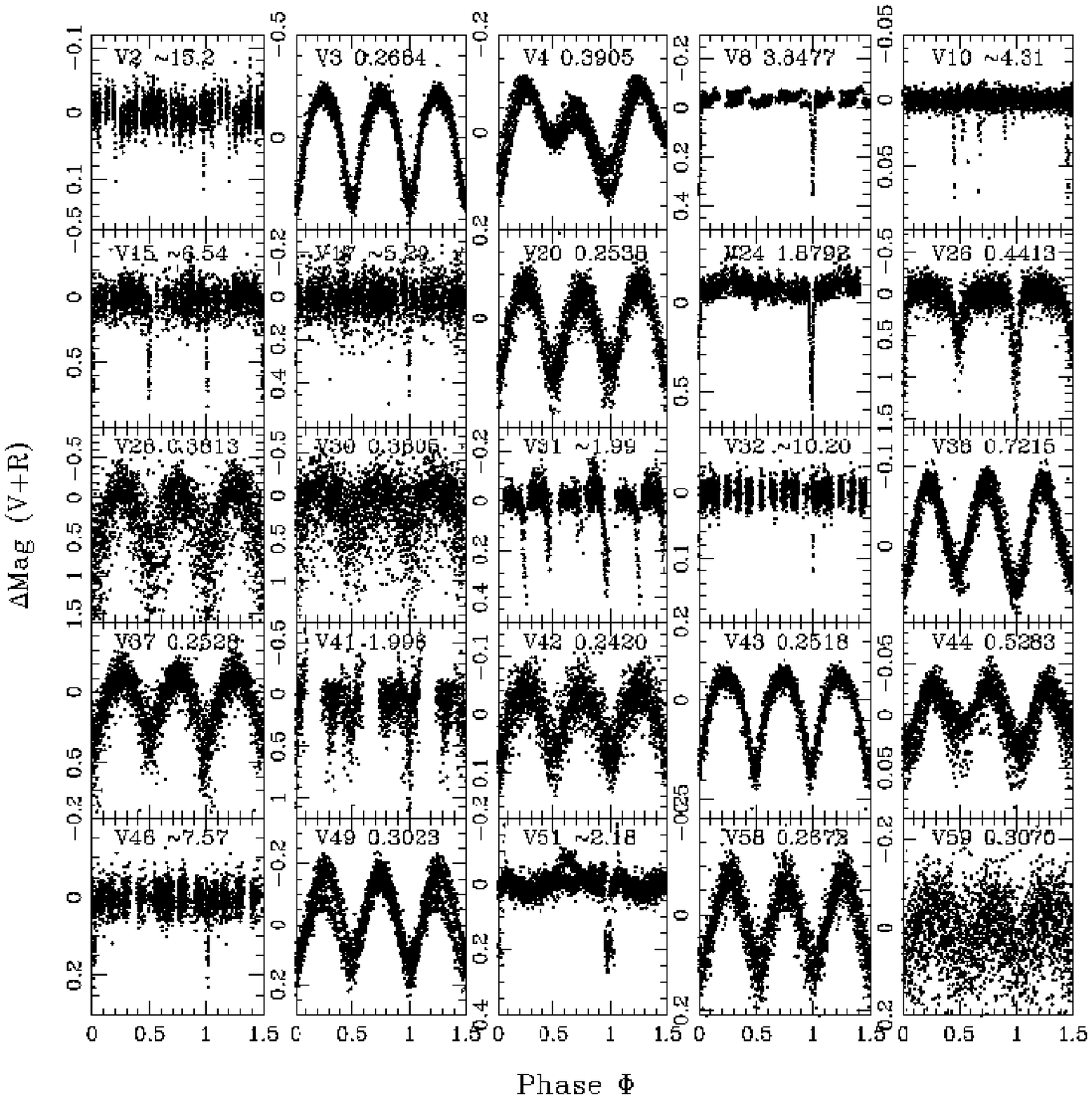} \figcaption[weldrake_fig8.eps]{Phase-wrapped V$+$R lightcurves for the detected EcB. The identification and period (in days) are overplotted for each system.\label{varplot1}}

\plotone{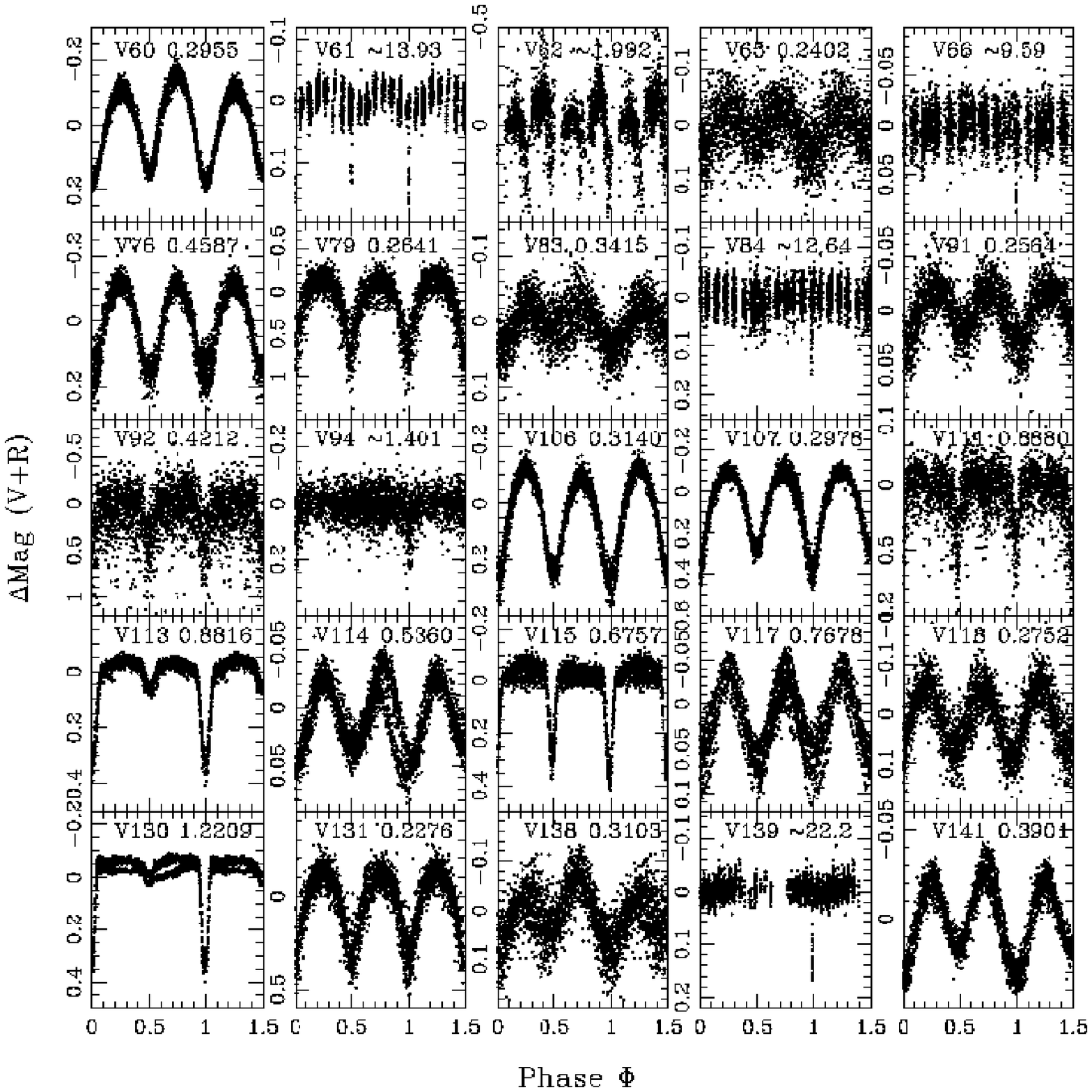} \figcaption[weldrake_fig9.eps]{Binary lightcurves
cont'd.\label{varplot2}}

\plotone{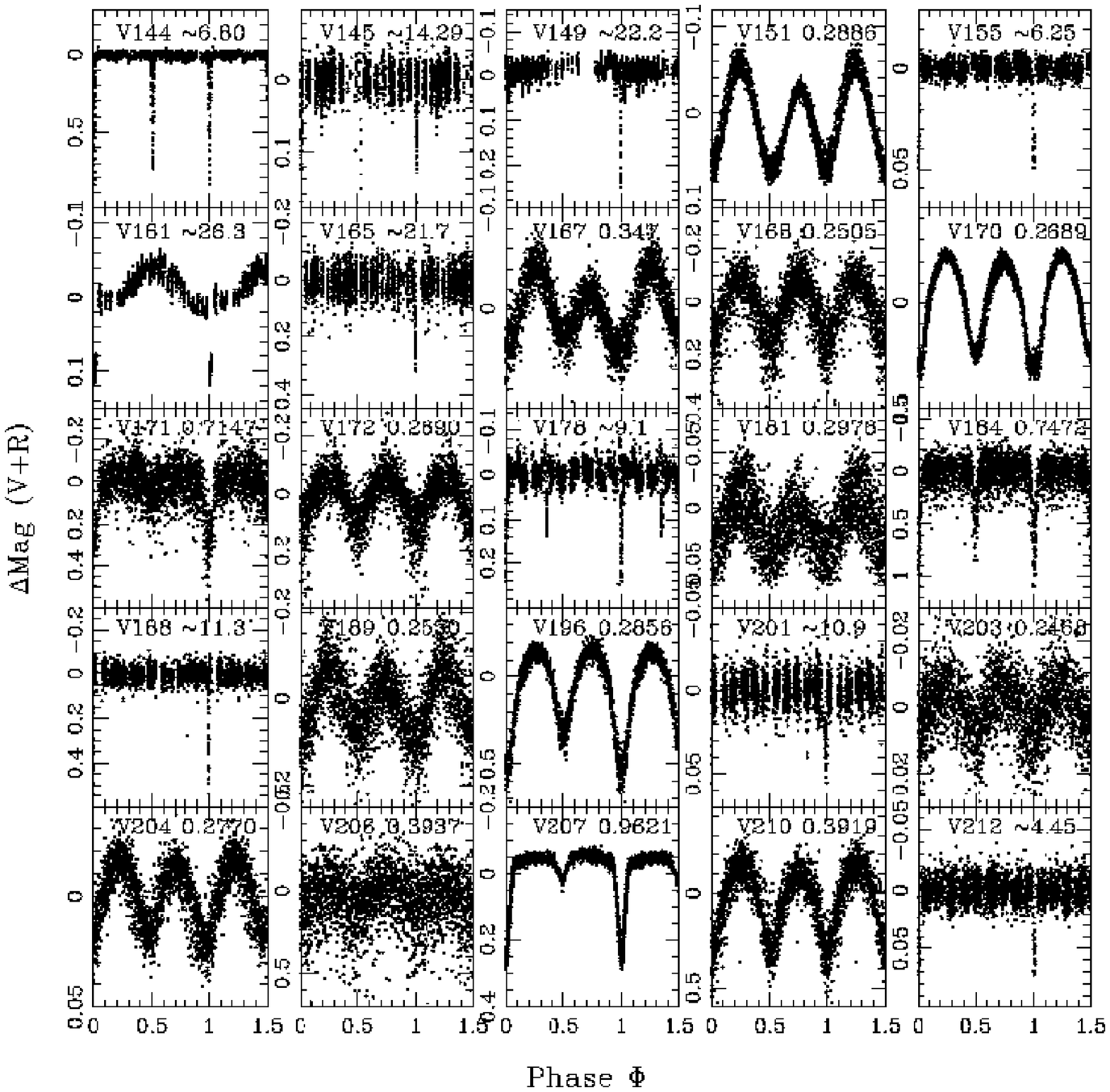} \figcaption[weldrake_fig10.eps]{Binary lightcurves
cont'd.\label{varplot3}}

\plotone{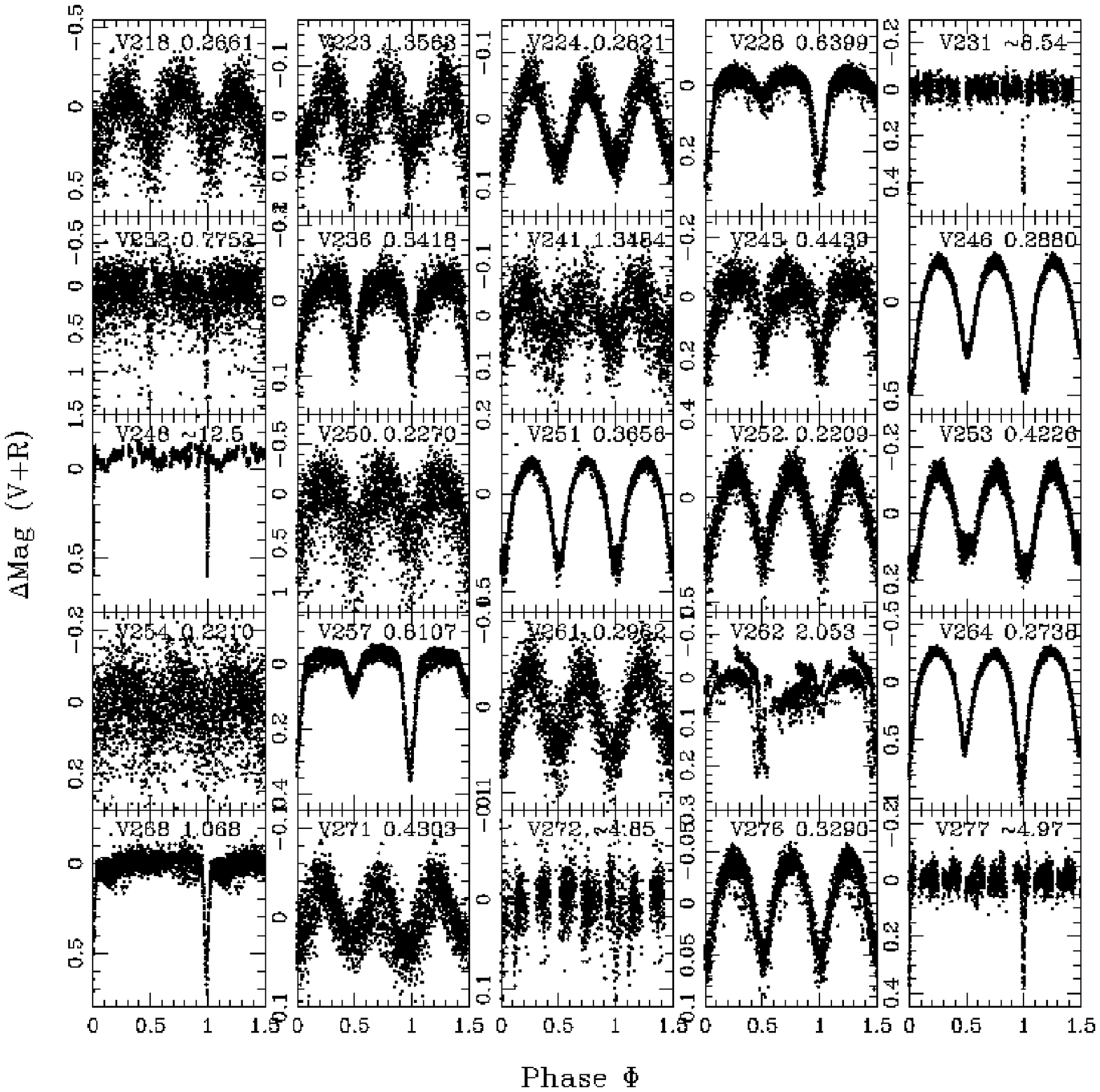} \figcaption[weldrake_fig11.eps]{Binary lightcurves
cont'd.\label{varplot4}}

\plotone{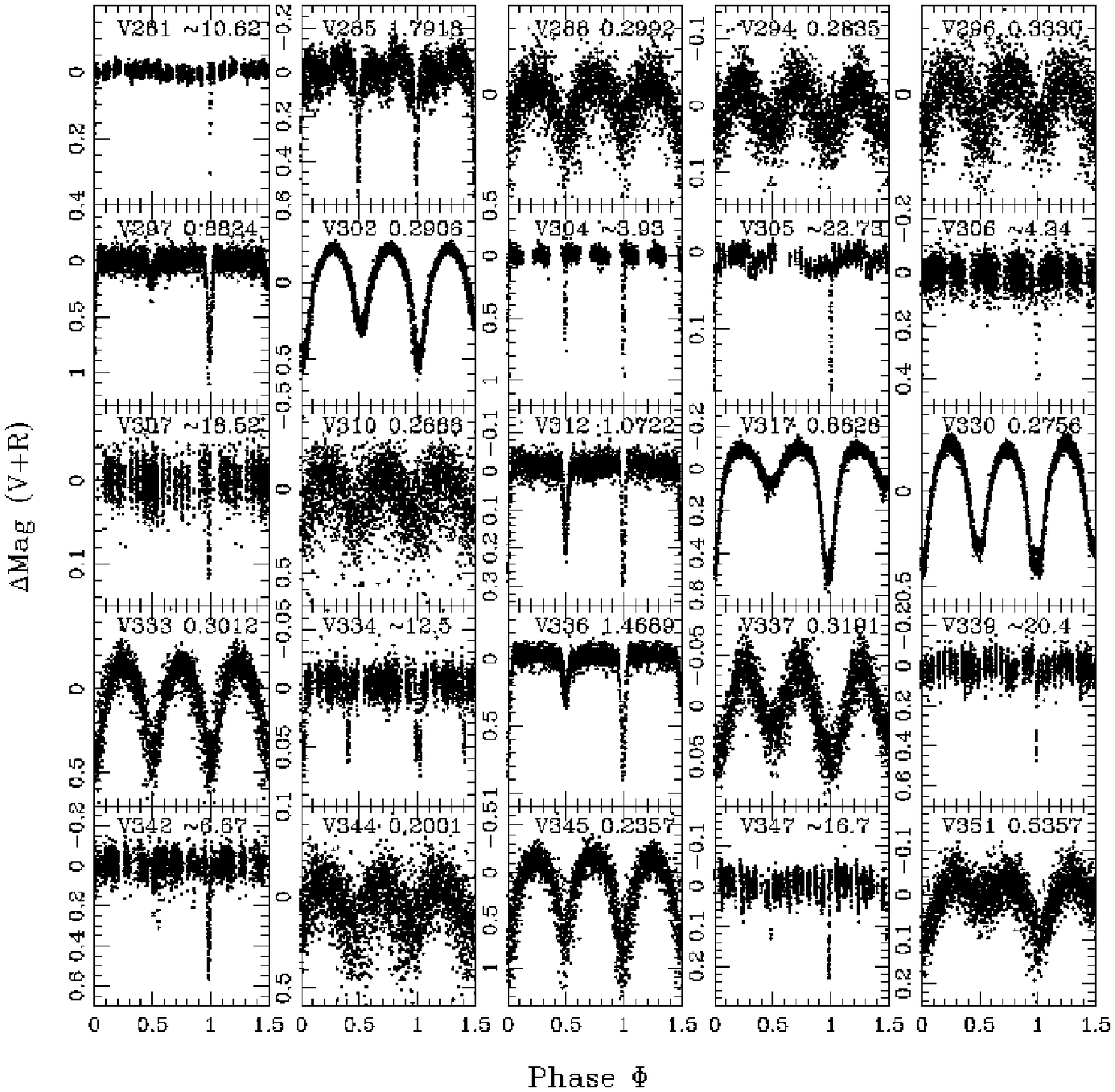} \figcaption[weldrake_fig12.eps]{Binary lightcurves
cont'd.\label{varplot5}}

\plotone{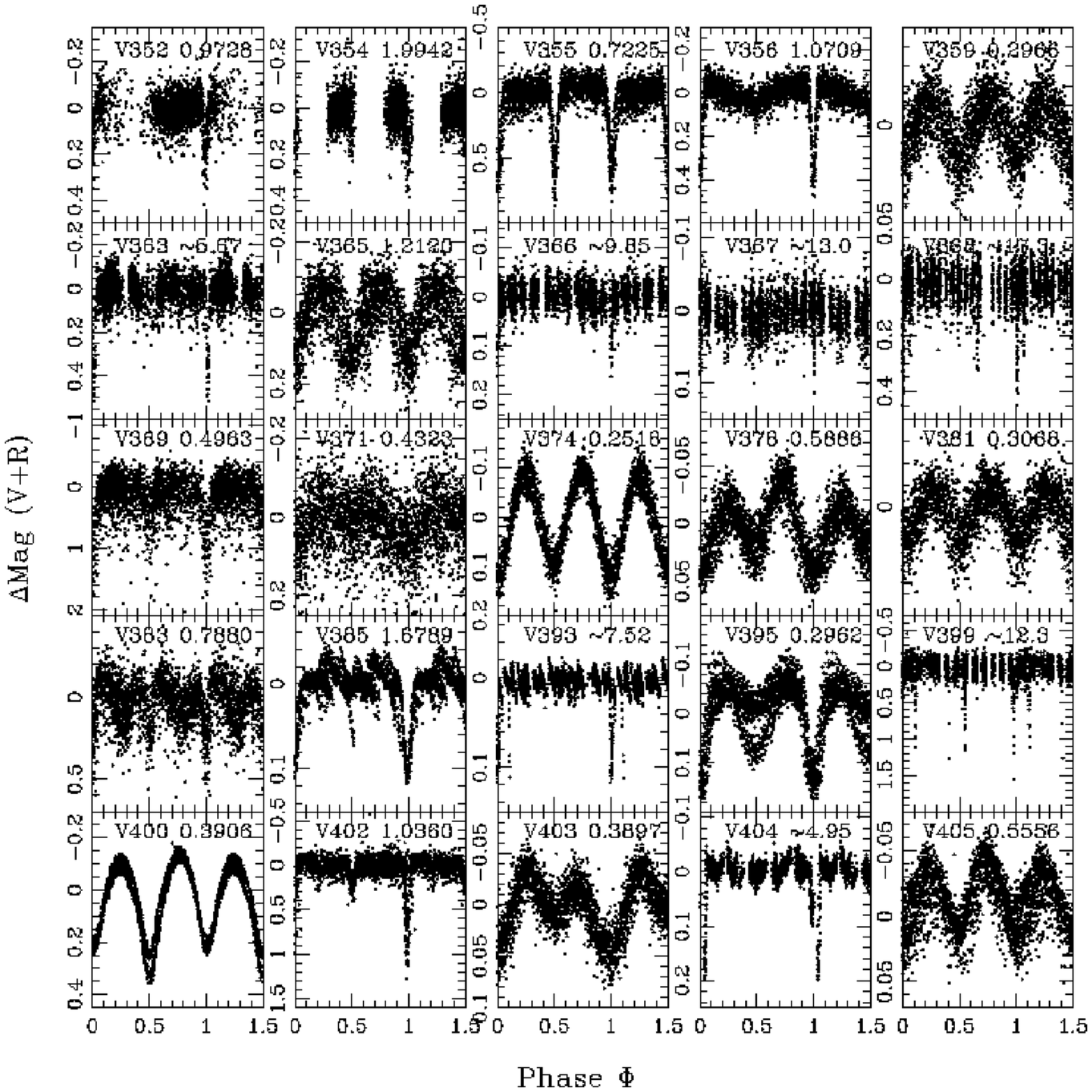} \figcaption[weldrake_fig13.eps]{Binary lightcurves
cont'd.\label{varplot6}}

\plotone{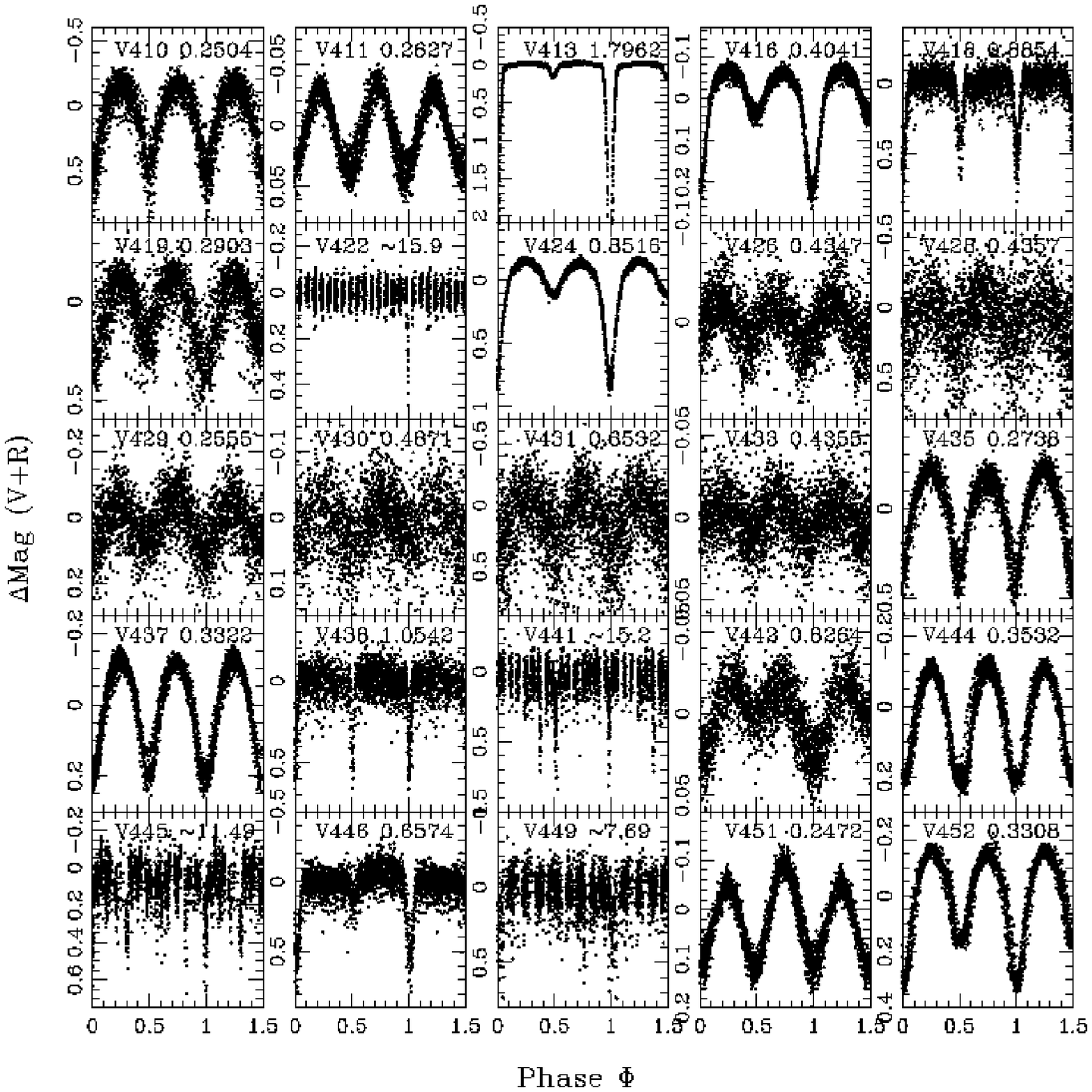} \figcaption[weldrake_fig14.eps]{Binary lightcurves
cont'd.\label{varplot7}}

\plotone{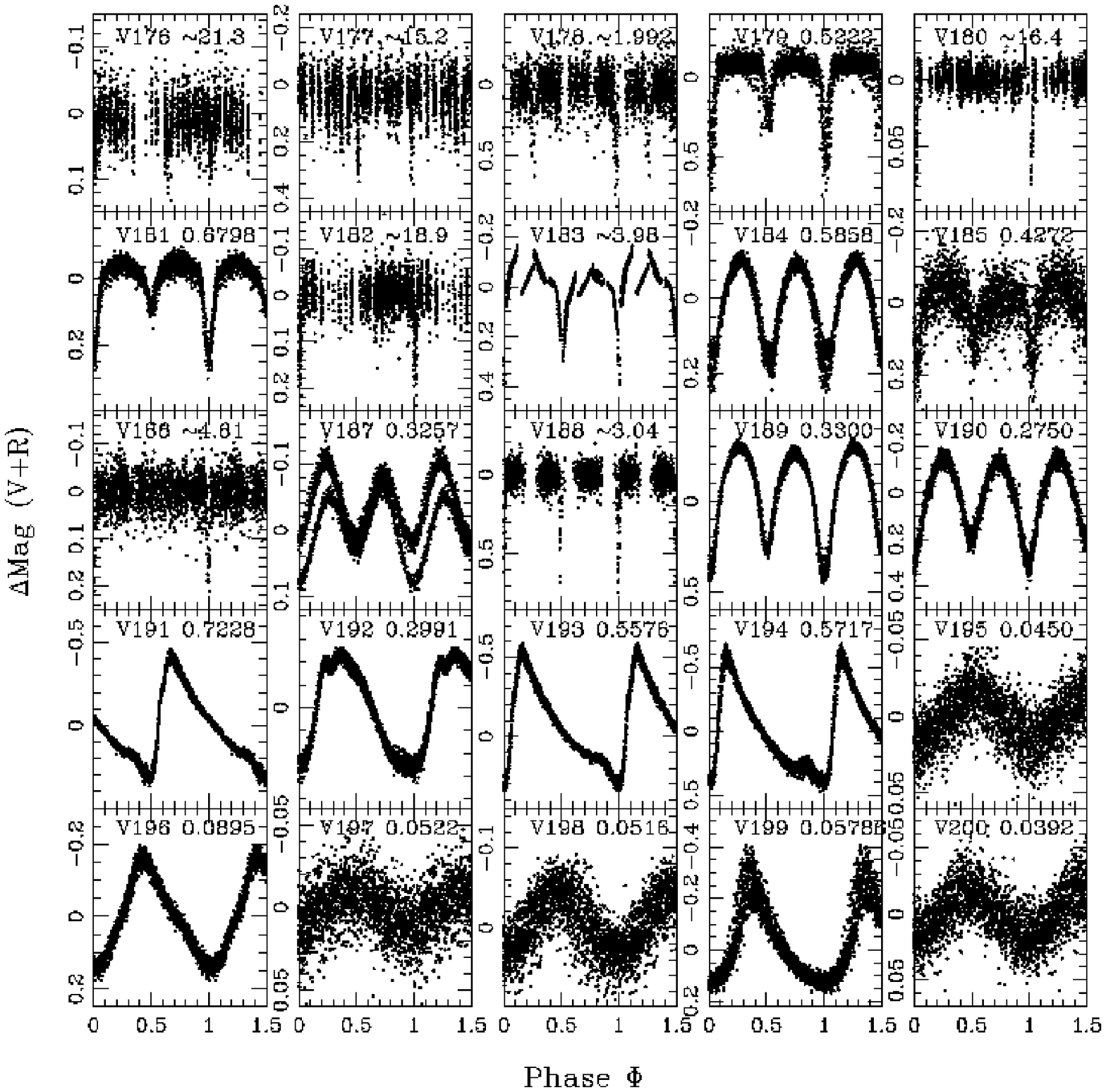} \figcaption[weldrake_fig15.eps]{Binary, RR Lyrae (V191+)
and $\delta$ Scuti (V195+) lightcurves.\label{varplot8}}

\plotone{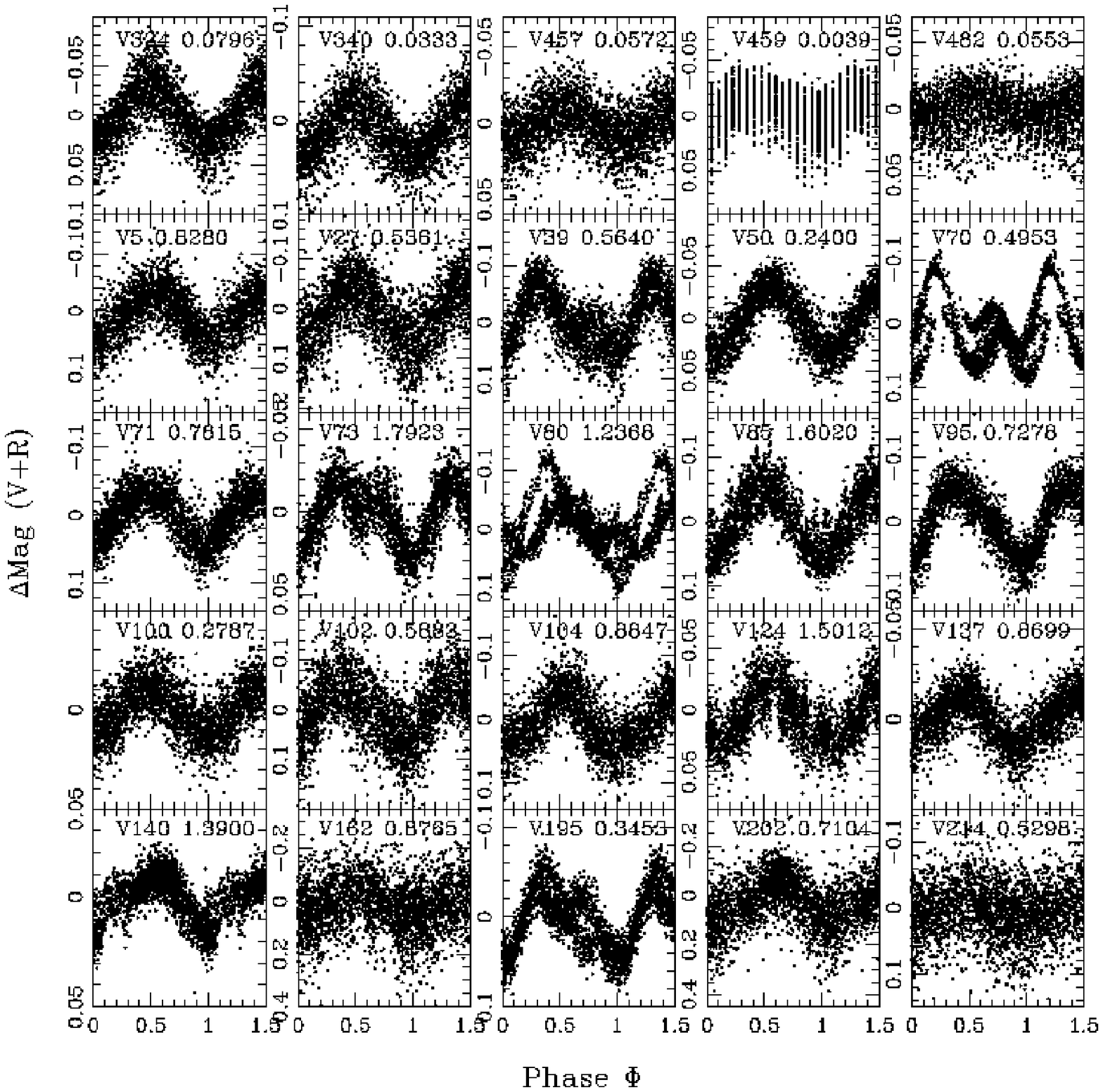} \figcaption[weldrake_fig16.eps]{$\delta$ Scuti
lightcurves cont'd and example pulsators (V5+).\label{varplot9}}

\plotone{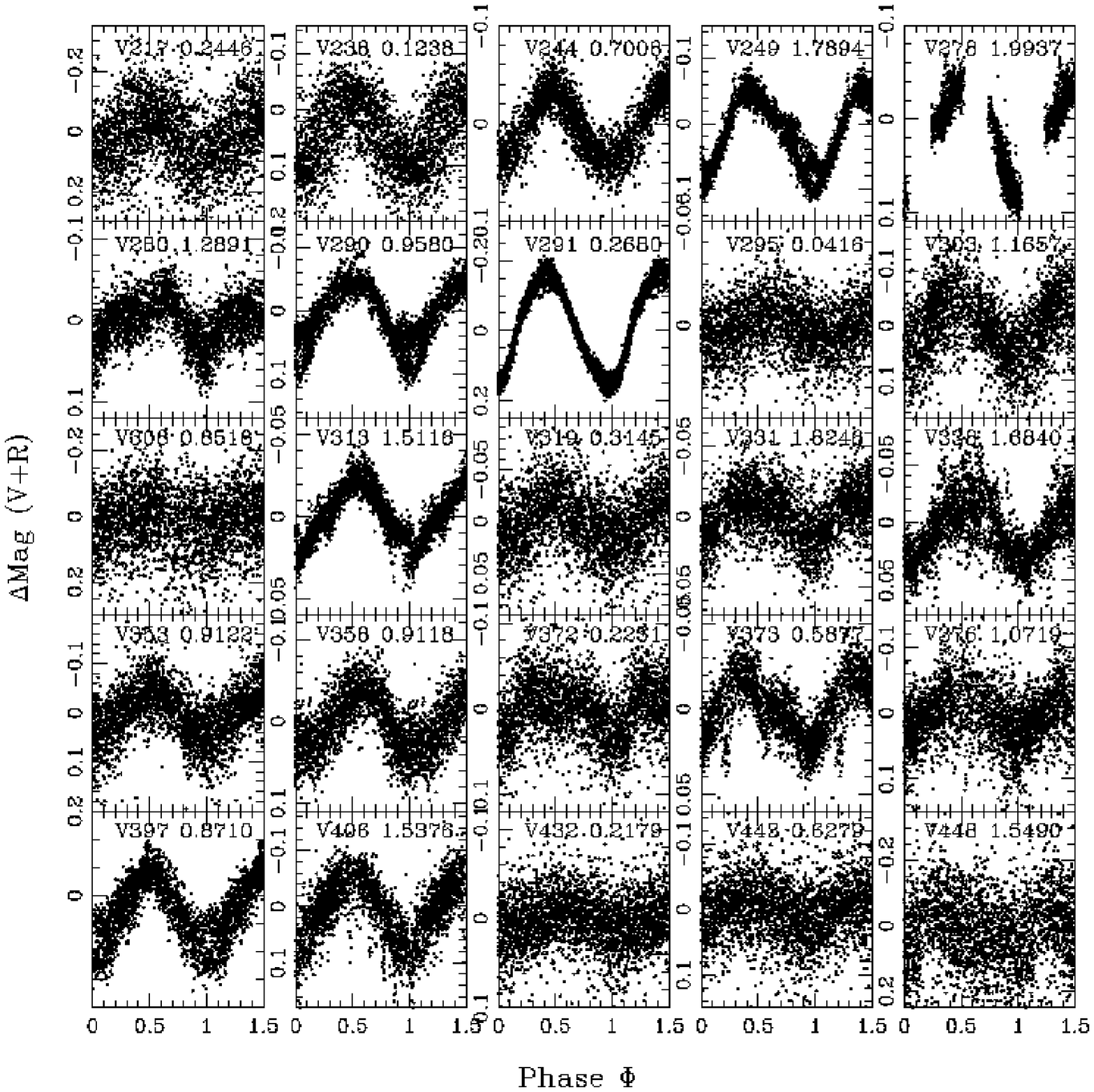} \figcaption[weldrake_fig17.eps]{Example pulsators
cont'd.\label{varplot10}}

\begin{center}
\centering \footnotesize
\begin{tabular}{lll} \hline
\noalign{\medskip} $CCD$ & $\it{RA}(J2000.0)$ & $\it{DEC}(J2000.0)$ \\
& h\hspace{2mm}m\hspace{2mm}s &
$^{\circ}$\hspace{5mm}$'$\hspace{3mm}$''$ \\ \noalign{\medskip} \hline
\noalign{\medskip} 
1 & 15:31:47 & $-$42:34:45 \\ 
2 & 15:31:49 & $-$42:47:35 \\ 
3 & 15:31:50 & $-$43:01:04 \\ 
4 & 15:31:51 & $-$43:13:57 \\ 
5 & 15:29:30 & $-$43:14:06 \\ 
6 & 15:29:29 & $-$43:01:07 \\ 
7 & $-$ & $-$ \\ 
8 & 15:29:27 & $-$42:35:17 \\
\noalign{\medskip} \hline\hline
\end{tabular}
\end{center}
\vspace*{3mm} \normalsize{Table 1: Equatorial coordinates (J2000.0)
for the centers of our CCDs. CCD7 was inoperative and is marked with a
`$-$'.}

\newpage

\centering \footnotesize
\begin{tabular}{lllllllll}
ID&Type&Period (d)&RA (j2000.0)&DEC
(j2000.0)&V&$\sigma$V&V-I&$\sigma$(V-I)\\ & & &
h\hspace{2mm}m\hspace{2mm}s &
$^{\circ}$\hspace{5mm}$'$\hspace{3mm}$''$ & & & & \\
\hline \hline V2 & EcB & $\sim$15.15 & 15:28:30.29 &-42:59:51.98 &
18.187 & 0.005 & 0.744 & 0.008\\ V3 & W UMa & 0.26837 & 15:28:30.98
&-43:00:17.23 & 18.830 & 0.006 & 0.779 & 0.008\\ V4 & EcB & 0.39051 &
15:28:31.49 &-43:07:35.66 & 16.699 & 0.004 & 0.819 & 0.006\\ V8 & EcB
& 3.84765 & 15:28:34.03 &-42:31:00.60 & 15.739 & 0.005 & 0.838 &
0.013\\ V10 & EcB & $\sim$4.31 & 15:28:35.46 &-42:33:17.75 & 15.860 &
0.003 & 0.660 & 0.007\\ V15 & EcB & $\sim$6.54 & 15:28:38.99
&-42:33:23.34 & 20.309 & 0.017 & 0.945 & 0.022\\ V17 & EcB & 5.2915 &
15:28:40.56 &-42:56:54.39 & 19.960 & 0.013 & 0.655 & 0.018\\ V20 & W
UMa & 0.25374 & 15:28:41.36 &-43:05:08.04 & 19.786 & 0.012 & 0.800 &
0.016\\ V24 & EcB & 1.87920 & 15:28:42.69 &-42:59:47.78 & 18.982 &
0.007 & 1.224 & 0.009\\ V26 & EcB & 0.44130 & 15:28:43.13
&-43:14:25.41 & 21.052 & 0.034 & 1.924 & 0.036\\ V28 & W UMa & 0.38132
& 15:28:43.58 &-42:34:28.64 & - & - & - & -\\ V30
& W UMa & 0.3805 & 15:28:44.60 &-42:34:37.89 & 21.667 & 0.066 & 0.977
& 0.084\\ V31 & EcB & $\sim$1.99 & 15:28:44.78 &-43:10:47.53 & 18.282
& 0.017 & 0.618 & 0.021\\ V32 & EcB & $\sim$10.20 & 15:28:45.73
&-42:40:10.78 & 17.850 & 0.003 & 0.843 & 0.004\\ V36 & EcB & 0.72150 &
15:28:47.82 &-42:58:03.82 & 16.657 & 0.002 & 0.414 & 0.003\\ V37 & W
UMa & 0.25278 & 15:28:48.94 &-42:33:14.88 & 20.181 & 0.022 & 0.960 &
0.026\\ V41 & EcB & 1.99634 & 15:28:51.92 &-42:37:53.74 & 20.943 &
0.073 & 1.406 & 0.093\\ V42 & W UMa & 0.24198 & 15:28:52.59
&-43:10:24.26 & 18.985 & 0.006 & 0.681 & 0.008\\ V43 & W UMa & 0.25179
& 15:28:53.83 &-42:34:08.44 & 19.185 & 0.009 & 0.936 & 0.011\\ V44 &
EcB & 0.5283 & 15:28:54.17 &-43:09:26.08 & 15.145 & 0.004 & 0.342 &
0.016\\ V46 & EcB & 7.574 & 15:28:56.99 &-43:05:43.42 & 20.307 & 0.014
& 0.670 & 0.023\\ V49 & W UMa & 0.30227 & 15:28:59.06 &-43:13:37.12 &
18.844 & 0.006 & 0.962 & 0.008\\ V51 & EcB & $\sim$2.18 & 15:29:00.20
&-43:03:58.98 & 18.437 & 0.005 & 1.068 & 0.007\\ V58 & W UMa & 0.26719
& 15:29:03.34 &-42:33:53.89 & 19.000 & 0.008 & 0.843 & 0.011\\ V59 & W
UMa & 0.30699 & 15:29:03.53 &-42:31:08.95 & 19.885 & 0.012 & 0.682 &
0.017\\ V60 & W UMa & 0.29546 & 15:29:05.20 &-43:07:41.87 & 15.832 &
0.004 & 0.807 & 0.012\\ V61 & EcB & 13.926 & 15:29:05.65 &-43:19:09.02
& 16.881 & 0.012 & 0.853 & 0.014\\ V62 & EcB & $\sim$1.99 &
15:29:06.26 &-43:10:21.02 & - & - & - & -\\ V65 &
W UMa & 0.24015 & 15:29:08.87 &-42:30:45.17 & 19.363 & 0.007 & 1.162 &
0.009\\ V66 & EcB & $\sim$9.59 & 15:29:09.23 &-42:32:49.89 & 18.779 &
0.007 & 2.204 & 0.008\\ V76 & EcB & 0.45872 & 15:29:14.77
&-42:38:36.87 & 18.226 & 0.008 & 0.454 & 0.013\\ V79 & W UMa & 0.26406
& 15:29:15.99 &-42:40:58.84 & 20.690 & 0.034 & 1.001 & 0.040\\ V83 & W
UMa & 0.34154 & 15:29:16.59 &-43:11:37.26 & 19.397 & 0.009 & 2.151 &
0.010\\ V84 & EcB & 12.64 & 15:29:16.61 &-42:59:34.17 & 18.876 & 0.006
& 0.688 & 0.008\\ V91 & W UMa & 0.2564 & 15:29:20.71 &-43:10:56.27 &
18.132 & 0.004 & 0.918 & 0.006\\ V92 & EcB & 0.42122 & 15:29:20.84
&-43:17:55.92 & 21.863 & 0.066 & 1.353 & 0.079\\ V94 & EcB &
$\sim$1.40 & 15:29:23.10 &-43:12:35.69 & 19.826 & 0.014 & 0.912 &
0.019\\ V106 & W UMa & 0.31400 & 15:29:30.08 &-42:41:12.99 & 18.162 &
0.005 & 0.712 & 0.007\\ V107 & W UMa & 0.29784 & 15:29:30.34
&-43:12:50.70 & 18.788 & 0.014 & 0.815 & 0.018\\ V111 & EcB & 0.88795
& 15:29:36.30 &-43:10:03.20 & 21.168 & 0.047 & 1.326 & 0.050\\ V113 &
EcB & 0.88159 & 15:29:38.56 &-42:56:55.11 & 16.024 & 0.002 & 0.346 &
0.005\\ V114 & EcB & 0.5360 & 15:29:39.23 &-43:09:59.61 & 16.688 &
0.003 & 0.730 & 0.005\\ V115 & EcB & 0.67566 & 15:29:39.31
&-43:13:31.04 & 18.595 & 0.006 & 0.860 & 0.007\\ V117 & EcB & 0.76777
& 15:29:40.83 &-42:38:03.16 & 15.519 & 0.003 & 0.480 & 0.005\\ V118 &
W UMa & 0.27518 & 15:29:41.11 &-42:56:30.49 & 19.029 & 0.013 & 0.848 &
0.016\\ V130 & EcB & 1.22086 & 15:29:44.70 &-43:02:48.30 & 16.128 &
0.002 & 0.718 & 0.005\\ V131 & W UMa & 0.22763 & 15:29:44.70
&-43:17:03.90 & 19.843 & 0.022 & 1.026 & 0.029\\ V138 & W UMa & 0.3103
& 15:29:48.59 &-43:02:52.47 & 19.061 & 0.013 & 0.701 & 0.022\\ \hline
\end{tabular}

\centering \footnotesize
\begin{tabular}{lllllllll}
ID&Type&Period (d)&RA (j2000.0)&DEC
(j2000.0)&V&$\sigma$V&V-I&$\sigma$(V-I)\\ & & &
h\hspace{2mm}m\hspace{2mm}s &
$^{\circ}$\hspace{5mm}$'$\hspace{3mm}$''$ & & & & \\
\hline \hline 
V139 &
EcB & $\sim$22.22 & 15:29:48.82 &-43:16:58.56 & 17.694 & 0.004 & 0.686
& 0.006\\ V141 & EcB & 0.39008 & 15:29:49.42 &-42:34:39.26 & 15.947 &
0.002 & 0.773 & 0.004\\ V144 & EcB & 6.8016 & 15:29:50.27
&-43:11:33.10 & 17.931 & 0.008 & 0.871 & 0.011\\ V145 & EcB &
$\sim$14.29 & 15:29:50.34 &-43:11:31.14 & 18.442 & 0.019 & 1.129 &
0.022\\ V149 & EcB & $\sim$22.22 & 15:29:52.08 &-43:17:48.50 & 17.251
& 0.003 & 0.672 & 0.006\\ V151 & W UMa & 0.28859 & 15:29:55.18
&-43:13:13.61 & 15.754 & 0.007 & 0.680 & 0.018\\ V155 & EcB & 6.2499 &
15:29:58.17 &-43:06:31.31 & 15.195 & 0.002 & - & -\\ V161 &
EcB & $\sim$26.29 & 15:30:02.45 &-43:00:18.92 & 16.394 & 0.003 & 0.955
& 0.009\\ V165 & EcB & $\sim$21.74 & 15:30:04.23 &-42:37:00.12 &
19.509 & 0.011 & 0.664 & 0.015\\ V167 & W UMa & 0.347 & 15:30:07.20
&-43:01:11.10 & 17.746 & 0.005 & 0.776 & 0.010\\ V168 & W UMa &
0.25049 & 15:30:07.91 &-42:33:38.39 & 19.756 & 0.010 & 0.402 & 0.020\\
V170 & W UMa & 0.26889 & 15:30:08.27 &-42:35:09.22 & 18.097 & 0.004 &
0.879 & 0.005\\ V171 & EcB & 0.71467 & 15:30:08.27 &-43:13:53.74 &
20.152 & 0.024 & 0.560 & 0.039\\ V172 & W UMa & 0.26898 & 15:30:08.48
&-42:35:11.50 & 19.649 & 0.016 & 1.061 & 0.018\\ V178 & EcB & 9.111 &
15:30:11.46 &-42:36:57.65 & 18.406 & 0.005 & 0.945 & 0.006\\ V181 & W
UMa & 0.29761 & 15:30:13.45 &-42:33:41.15 & 17.322 & 0.003 & 1.226 &
0.004\\ V184 & EcB & 0.74719 & 15:30:14.41 &-43:12:45.97 & 21.448 &
0.047 & 2.602 & 0.048\\ V188 & EcB & $\sim$11.30 & 15:30:16.82
&-42:33:04.76 & 18.844 & 0.007 & 0.692 & 0.009\\ V189 & W UMa &
0.25502 & 15:30:17.21 &-43:15:34.25 & 19.824 & 0.011 & 0.961 & 0.014\\
V196 & W UMa & 0.28558 & 15:30:18.40 &-42:37:55.20 & 19.129 & 0.008 &
0.963 & 0.011\\ V201 & EcB & $\sim$10.87 & 15:30:20.94 &-43:20:07.91 &
16.907 & 0.007 & 0.613 & 0.012\\ V203 & W UMa & 0.24680 & 15:30:21.55
&-42:37:22.49 & 17.151 & 0.004 & 0.721 & 0.009\\ V204 & W UMa &
0.27700 & 15:30:26.72 &-42:40:42.60 & 16.519 & 0.004 & 0.989 & 0.007\\
V206 & EcB & 0.39372 & 15:30:28.20 &-42:30:53.15 & 21.036 & 0.038 &
1.075 & 0.046\\ V207 & EcB & 0.96208 & 15:30:28.37 &-43:09:15.62 &
16.850 & 0.016 & 0.658 & 0.020\\ V210 & EcB & 0.39185 & 15:30:30.62
&-43:20:34.62 & 19.809 & 0.015 & 0.919 & 0.018\\ 
V212 & EcB & $\sim$4.45 & 15:30:31.41 &-43:13:50.00 &
17.515 & 0.005 & 0.670 & 0.007\\ V218 & W UMa & 0.26605 & 15:30:33.91
&-43:01:35.66 & 20.762 & 0.031 & 0.728 & 0.043\\ V223 & EcB & 1.35629
& 15:30:34.85 &-43:10:43.63 & 19.521 & 0.012 & 1.322 & 0.014\\ V224 &
W UMa & 0.28208 & 15:30:36.66 &-43:18:22.33 & 17.349 & 0.005 & 1.172 &
0.006\\ V228 & EcB & 0.63984 & 15:30:49.60 &-43:00:30.86 & 15.553 &
0.003 & 0.455 & 0.005\\ V231 & EcB & 8.5470 & 15:30:52.24
&-42:51:39.28 & 18.527 & 0.009 & 0.784 & 0.013\\ V232 & EcB & 0.77521
& 15:30:52.28 &-42:32:53.14 & 21.985 & 0.066 & 2.910 & 0.067\\ V236 &
W UMa & 0.34175 & 15:30:53.68 &-42:36:16.14 & 18.361 & 0.004 & 2.028 &
0.006\\ V241 & EcB & 1.3484 & 15:30:56.15 &-42:50:40.36 & 19.227 &
0.010 & 1.343 & 0.012\\ V243 & EcB & 0.44388 & 15:30:57.35
&-42:47:51.39 & 19.304 & 0.013 & 1.502 & 0.014\\ V246 & W UMa &
0.28803 & 15:30:59.40 &-43:12:28.72 & 17.280 & 0.003 & 0.777 & 0.005\\
V248 & EcB & $\sim$12.50 & 15:31:00.46 &-42:53:46.96 & 17.612 & 0.007
& 1.060 & 0.009\\ V250 & W UMa & 0.22700 & 15:31:01.31 &-42:52:32.17 &
21.226 & 0.053 & 0.820 & 0.070\\ V251 & W UMa & 0.36560 & 15:31:01.84
&-42:59:14.87 & 17.627 & 0.025 & 0.635 & 0.037\\ V252 & W UMa &
0.22092 & 15:31:01.99 &-42:32:30.18 & 19.632 & 0.012 & 1.138 & 0.016\\
V253 & EcB & 0.42261 & 15:31:02.16 &-43:13:15.08 & 17.906 & 0.003 &
0.370 & 0.006\\ V254 & W UMa & 0.22100 & 15:31:02.24 &-42:32:31.10 &
20.362 & 0.020 & 1.378 & 0.026\\ V257 & EcB & 0.81069 & 15:31:02.97
&-42:53:23.87 & 15.086 & 0.007 & 0.437 & 0.011\\ V261 & W UMa &
0.29620 & 15:31:04.04 &-43:09:36.05 & 18.871 & 0.005 & 0.680 & 0.007\\
V262 & EcB & $\sim$2.05 & 15:31:04.77 &-42:57:10.27 & 16.163 & 0.010 &
0.629 & 0.016\\ V264 & W UMa & 0.27379 & 15:31:05.24 &-42:47:35.15 &
17.664 & 0.007 & 1.136 & 0.009\\ V268 & EcB & 1.06799 & 15:31:06.41
&-42:33:47.29 & 18.914 & 0.006 & 1.187 & 0.008\\ V271 & EcB & 0.43028
& 15:31:07.11 &-42:57:47.33 & 18.379 & 0.008 & 0.989 & 0.011\\ \hline
\end{tabular}

\centering \footnotesize
\begin{tabular}{lllllllll}
ID&Type&Period (d)&RA (j2000.0)&DEC
(j2000.0)&V&$\sigma$V&V-I&$\sigma$(V-I)\\ & & &
h\hspace{2mm}m\hspace{2mm}s &
$^{\circ}$\hspace{5mm}$'$\hspace{3mm}$''$ & & & & \\
\hline \hline 
V272 &
EcB & 4.85015 & 15:31:07.73 &-42:32:57.29 & 17.920 & 0.014 & 0.665 &
0.018\\ V276 & W UMa & 0.32895 & 15:31:09.15 &-43:03:25.33 & 15.755 &
0.007 & 0.679 & 0.025\\ V277 & EcB & $\sim$4.97 & 15:31:09.15
&-43:15:48.63 & 18.812 & 0.006 & 0.713 & 0.009\\ V281 & EcB &
$\sim$10.62 & 15:31:11.04 &-43:09:16.77 & 17.403 & 0.020 & 0.743 &
0.030\\ V285 & EcB & 1.79179 & 15:31:13.33 &-42:39:28.98 & 18.778 &
0.011 & 0.640 & 0.016\\ V288 & W UMa & 0.29922 & 15:31:13.99
&-43:16:54.92 & 20.493 & 0.026 & 0.710 & 0.047\\ V294 & W UMa &
0.28345 & 15:31:18.01 &-43:15:55.56 & 18.989 & 0.012 & 0.463 & 0.015\\
V296 & W UMa & 0.33300 & 15:31:18.78 &-42:56:52.86 & 20.567 & 0.023 &
0.384 & 0.034\\ V297 & EcB & 0.88242 & 15:31:19.61 &-42:44:36.06 &
19.967 & 0.019 & 1.560 & 0.021\\ V302 & W UMa & 0.29058 & 15:31:20.66
&-42:58:33.50 & 17.987 & 0.004 & 0.916 & 0.006\\ V304 & EcB & 3.9356 &
15:31:21.14 &-42:34:28.85 & 19.264 & 0.008 & 2.397 & 0.009\\ V305 &
EcB & $\sim$22.73 & 15:31:21.81 &-43:16:22.69 & 16.200 & 0.002 & 0.690
& 0.005\\ V306 & EcB & $\sim$4.24 & 15:31:22.09 &-43:15:13.20 & 19.720
& 0.010 & 0.886 & 0.013\\ V307 & EcB & $\sim$18.52 & 15:31:23.22
&-43:02:24.95 & 18.695 & 0.004 & 0.618 & 0.008\\ V310 & W UMa &
0.26656 & 15:31:23.77 &-42:47:17.32 & 20.415 & 0.028 & 0.838 & 0.036\\
V312 & EcB & 1.07219 & 15:31:24.33 &-42:44:50.66 & 18.209 & 0.008 &
1.241 & 0.011\\ V317 & EcB & 0.86278 & 15:31:25.92 &-43:00:49.14 &
18.680 & 0.006 & 0.439 & 0.009\\ V330 & W UMa & 0.27555 & 15:31:34.44
&-43:12:59.60 & 18.229 & 0.004 & 0.817 & 0.007\\ V333 & W UMa &
0.30121 & 15:31:36.09 &-43:11:45.10 & 19.721 & 0.016 & 0.650 & 0.021\\
V334 & EcB & $\sim$12.50 & 15:31:37.13 &-43:10:16.74 & 17.575 & 0.003
& 0.662 & 0.005\\ V336 & EcB & 1.46889 & 15:31:37.47 &-42:56:07.74 &
19.629 & 0.009 & 0.754 & 0.011\\ V337 & W UMa & 0.3191 & 15:31:37.48
&-43:06:45.02 & 18.421 & 0.005 & 0.768 & 0.008\\ V339 & EcB &
$\sim$20.41 & 15:31:37.79 &-42:52:30.58 & 19.003 & 0.052 & 0.887 &
0.081\\ V342 & EcB & 6.6675 & 15:31:38.49 &-43:06:52.80 & 19.933 &
0.014 & 0.583 & 0.019\\ V344 & W UMa & 0.2009 & 15:31:41.31
&-43:12:17.09 & 20.925 & 0.027 & 1.014 & 0.034\\ V345 & W UMa &
0.23572 & 15:31:41.36 &-42:47:33.09 & 20.404 & 0.036 & 1.321 & 0.039\\
V347 & EcB & $\sim$16.67 & 15:31:42.08 &-43:11:04.15 & 18.514 & 0.006
& 0.722 & 0.008\\ V351 & EcB & 0.53566 & 15:31:44.85 &-42:47:44.64 &
18.264 & 0.009 & 0.794 & 0.014\\ V352 & EcB & 0.97276 & 15:31:45.72
&-42:53:13.03 & 19.853 & 0.021 & 1.855 & 0.022\\ V354 & EcB & 1.99421
& 15:31:45.98 &-42:29:26.55 & 19.478 & 0.012 & 1.088 & 0.019\\ V355 &
EcB & 0.7225 & 15:31:46.46 &-42:45:05.96 & 20.126 & 0.019 & 1.605 &
0.023\\ V356 & EcB & 1.07088 & 15:31:46.49 &-42:52:23.93 & 19.180 &
0.012 & 1.392 & 0.014\\ V359 & W UMa & 0.29655 & 15:31:46.85
&-43:06:06.24 & 18.226 & 0.010 & 1.388 & 0.012\\ V363 & EcB &
$\sim$5.67 & 15:31:49.28 &-43:08:41.05 & 19.923 & 0.011 & 0.675 &
0.015\\ V365 & EcB & 1.21198 & 15:31:50.64 &-43:19:08.61 & 19.804 &
0.013 & 1.530 & 0.015\\ V366 & EcB & $\sim$9.85 & 15:31:50.90
&-43:16:17.33 & 18.657 & 0.011 & 0.619 & 0.014\\ V367 & EcB &
$\sim$12.99 & 15:31:51.29 &-43:09:04.00 & 19.024 & 0.006 & 1.307 &
0.007\\ V368 & EcB & $\sim$17.25 & 15:31:51.56 &-42:31:39.91 & 19.568
& 0.007 & 0.694 & 0.015\\ V369 & EcB & 0.49627 & 15:31:52.01
&-42:47:27.83 & 21.507 & 0.056 & 1.279 & 0.068\\ V371 & EcB & 0.43234
& 15:31:52.65 &-42:55:30.38 & 20.113 & 0.017 & 1.434 & 0.018\\ V374 &
W UMa & 0.25177 & 15:31:54.55 &-42:54:06.47 & 18.183 & 0.006 & 1.200 &
0.012\\ V378 & EcB & 0.58859 & 15:31:55.98 &-43:04:22.07 & 17.753 &
0.004 & 0.930 & 0.006\\ V381 & W UMa & 0.30676 & 15:31:56.81
&-42:51:41.51 & 19.804 & 0.014 & 0.559 & 0.022\\ V383 & EcB & 0.78798
& 15:31:57.82 &-42:51:06.44 & 20.390 & 0.024 & 1.129 & 0.028\\ V385 &
EcB & 1.67889 & 15:31:59.34 &-43:18:09.60 & 15.514 & 0.003 & 0.244 &
0.006\\ V393 & EcB & $\sim$7.52 & 15:32:05.09 &-43:07:08.12 & 16.622 &
0.004 & 0.676 & 0.007\\ V395 & W UMa & 0.29619 & 15:32:05.78
&-42:42:36.44 & 17.483 & 0.004 & 1.365 & 0.007\\ V399 & EcB &
$\sim$12.35 & 15:32:06.93 &-43:03:03.05 & 20.896 & 0.025 & 1.219 &
0.035\\ V400 & EcB & 0.39061 & 15:32:06.95 &-43:19:09.01 & 16.715 &
0.006 & 0.641 & 0.010\\ \hline
\end{tabular}

\centering \footnotesize
\begin{tabular}{lllllllll}
ID&Type&Period (d)&RA (j2000.0)&DEC
(j2000.0)&V&$\sigma$V&V-I&$\sigma$(V-I)\\ & & &
h\hspace{2mm}m\hspace{2mm}s &
$^{\circ}$\hspace{5mm}$'$\hspace{3mm}$''$ & & & & \\
\hline \hline 
V402 & EcB & 1.03601 & 15:32:08.91
&-42:49:59.62 & 20.448 & 0.025 & 2.078 & 0.027\\ V403 & EcB & 0.3897 &
15:32:10.49 &-43:00:22.49 & 17.957 & 0.008 & 0.781 & 0.009\\ 
V404 & EcB & $\sim$4.95 & 15:32:10.70 &-43:18:01.36 &
17.529 & 0.003 & 0.685 & 0.004\\ V405 & EcB & 0.55556 & 15:32:10.87
&-43:09:09.48 & 16.694 & 0.002 & 0.272 & 0.005\\ V410 & W UMa &
0.25039 & 15:32:12.42 &-42:53:43.14 & 20.217 & 0.022 & 1.242 & 0.025\\
V411 & W UMa & 0.26266 & 15:32:12.96 &-42:38:31.50 & 16.143 & 0.003 &
0.748 & 0.004\\ V413 & EcB & 1.7962 & 15:32:15.56 &-42:43:21.56 &
14.595 & 0.008 & - & -\\ V416 & EcB & 0.40409 & 15:32:17.81
&-43:06:37.05 & 17.136 & 0.005 & 0.513 & 0.008\\ V418 & EcB & 0.88544
& 15:32:18.21 &-43:12:10.65 & 20.479 & 0.022 & 1.337 & 0.025\\ V419 &
W UMa & 0.29027 & 15:32:18.32 &-42:32:01.71 & 19.874 & 0.050 & 0.971 &
0.051\\ V422 & EcB & 15.8725 & 15:32:20.04 &-42:48:03.65 & 19.016 &
0.011 & 0.852 & 0.013\\ V424 & EcB & 0.85157 & 15:32:20.95
&-43:11:56.56 & 18.178 & 0.004 & 0.554 & 0.006\\ V426 & EcB & 0.43466
& 15:32:22.80 &-42:31:04.24 & 18.432 & 0.006 & 0.861 & 0.009\\ V428 &
EcB & 0.43572 & 15:32:23.27 &-42:31:16.79 & 21.520 & 0.048 & 1.806 &
0.055\\ V429 & W UMa & 0.2555 & 15:32:23.43 &-43:14:42.79 & 20.181 &
0.022 & 0.841 & 0.030\\ V430 & EcB & 0.48709 & 15:32:23.66
&-42:49:05.10 & 19.107 & 0.012 & 0.630 & 0.016\\ V431 & EcB & 0.65315
& 15:32:23.77 &-42:30:51.84 & 21.469 & 0.057 & 0.823 & 0.073\\ V433 &
EcB & 0.43546 & 15:32:24.83 &-42:31:09.56 & 17.888 & 0.003 & 0.591 &
0.005\\ V435 & W UMa & 0.27377 & 15:32:25.52 &-43:09:23.93 & 19.778 &
0.010 & 0.818 & 0.013\\ V437 & W UMa & 0.33218 & 15:32:27.21
&-42:34:14.80 & 17.783 & 0.003 & 0.662 & 0.006\\ V438 & EcB & 1.05423
& 15:32:28.48 &-43:16:09.16 & 20.326 & 0.040 & 1.876 & 0.044\\ V441 &
EcB & $\sim$15.15 & 15:32:28.89 &-43:08:58.22 & 20.242 & 0.019 & 0.810
& 0.022\\ V442 & EcB & 0.8264 & 15:32:28.90 &-42:57:08.37 & 17.588 &
0.006 & 0.498 & 0.009\\ V444 & W UMa & 0.35318 & 15:32:29.32
&-43:12:02.87 & 17.360 & 0.003 & 0.630 & 0.007\\ V445 & EcB &
$\sim$11.49 & 15:32:30.19 &-43:08:59.26 & 20.147 & 0.015 & 1.191 &
0.018\\ V446 & EcB & 0.65738 & 15:32:30.51 &-43:06:51.95 & 20.223 &
0.019 & 1.137 & 0.022\\ V449 & EcB & $\sim$7.69 & 15:32:32.17
&-43:08:53.59 & 20.544 & 0.020 & 0.580 & 0.028\\ V451 & W UMa &
0.24724 & 15:32:34.48 &-42:57:09.15 & 18.203 & 0.005 & 0.941 & 0.006\\
V452 & W UMa & 0.33083 & 15:32:34.63 &-42:51:05.65 & 17.443 & 0.011 &
0.611 & 0.029\\ V456 & EcB & $\sim$21.28 & 15:32:35.44 &-43:08:53.39 &
19.110 & 0.006 & 1.352 & 0.008\\ V458 & EcB & $\sim$15.15 &
15:32:36.48 &-43:09:59.90 & 19.770 & 0.009 & 1.006 & 0.012\\ V461 &
EcB & $\sim$1.99 & 15:32:37.22 &-42:58:25.87 & 20.524 & 0.040 & 1.099
& 0.052\\ V462 & EcB & 0.52219 & 15:32:37.32 &-42:48:19.15 & 19.150 &
0.011 & 1.630 & 0.012\\ V464 & EcB & $\sim$16.39 & 15:32:39.33
&-43:10:51.98 & 17.055 & 0.003 & 0.792 & 0.006\\ V467 & EcB & 0.67979
& 15:32:39.68 &-43:12:29.51 & 17.295 & 0.003 & 0.253 & 0.004\\ V469 &
EcB & $\sim$18.87 & 15:32:42.66 &-43:04:24.40 & 19.446 & 0.007 & 1.226
& 0.009\\ V472 & EcB & $\sim$3.98 & 15:32:45.16 &-43:15:47.04 & 14.766
& 0.010 & - & -\\ V476 & EcB & 0.58581 & 15:32:47.97
&-43:05:49.23 & 16.854 & 0.015 & 0.466 & 0.027\\ V477 & EcB & 0.42723
& 15:32:48.11 &-43:04:12.88 & 19.915 & 0.014 & 1.218 & 0.022\\ V480 &
EcB & 4.60817 & 15:32:49.05 &-42:33:40.34 & 19.025 & 0.006 & 0.869 &
0.010\\ V481 & W UMa & 0.32570 & 15:32:49.31 &-43:18:29.81 & 16.516 &
0.004 & 0.694 & 0.009\\ V483 & EcB & 3.04153 & 15:32:51.00
&-42:43:25.24 & 19.447 & 0.014 & 0.946 & 0.017\\ V492 & W UMa &
0.32995 & 15:32:58.43 &-43:05:21.34 & 16.495 & 0.003 & 0.793 & 0.006\\
V494 & W UMa & 0.27502 & 15:33:00.05 &-43:14:27.23 & 18.767 & 0.007 &
0.926 & 0.010\\ V29 & `AB' RR Lyr & 0.72284 & 15:28:44.53 &-43:11:26.93
& 15.202 & 0.003 & 0.432 & 0.011\\ V48 & `C' RR Lyr & 0.29914 &
15:28:58.14 &-43:11:59.12 & 15.663 & 0.002 & 0.221 & 0.005\\ V247 &
`AB' RR Lyr & 0.55755 & 15:31:00.05 &-43:13:53.02 & 15.507 & 0.002 &
0.503 & 0.013\\ V387 & `AB' RR Lyr & 0.57167 & 15:32:00.03
&-43:18:41.70 & 17.746 & 0.005 & 0.447 & 0.007\\ V221 & $\delta$ Scuti
& 0.044967 & 15:30:34.59 &-43:06:53.39 & 17.413 & 0.003 & 0.318 &
0.006\\ \hline
\end{tabular}

\centering \footnotesize
\begin{tabular}{lllllllll}
ID&Type&Period (d)&RA (j2000.0)&DEC
(j2000.0)&V&$\sigma$V&V-I&$\sigma$(V-I)\\ & & &
h\hspace{2mm}m\hspace{2mm}s &
$^{\circ}$\hspace{5mm}$'$\hspace{3mm}$''$ & & & & \\
\hline \hline 
V270 & $\delta$ Scuti & 0.089530 & 15:31:07.03 &-43:12:07.56 &
16.615 & 0.005 & 0.289 & 0.011\\ V283 & $\delta$ Scuti & 0.052169 &
15:31:12.45 &-43:01:49.96 & 18.045 & 0.004 & 0.282 & 0.007\\ V284 &
$\delta$ Scuti & 0.051600 & 15:31:12.82 &-42:40:10.26 & 17.065 & 0.003
& 0.343 & 0.005\\ V293 & $\delta$ Scuti & 0.057857 & 15:31:16.88
&-42:33:54.16 & 15.974 & 0.002 & 0.304 & 0.004\\ V322 & $\delta$ Scuti
& 0.039243 & 15:31:28.95 &-42:48:36.66 & 17.665 & 0.007 & 0.555 &
0.014\\ V324 & $\delta$ Scuti & 0.079601 & 15:31:29.54 &-43:00:00.51 &
17.416 & 0.004 & 0.305 & 0.006\\ V340 & $\delta$ Scuti & 0.033293 &
15:31:37.84 &-43:17:28.86 & 17.376 & 0.008 & 0.263 & 0.014\\ V457 &
$\delta$ Scuti & 0.057240 & 15:32:35.72 &-42:55:55.97 & 17.730 & 0.005
& 0.336 & 0.006\\ V459 & $\delta$ Scuti & 0.003919 & 15:32:36.67
&-42:32:21.98 & 16.017 & 0.002 & 0.427 & 0.003\\ V482 & $\delta$ Scuti
& 0.055329 & 15:32:49.79 &-42:36:31.65 & 16.082 & 0.003 & 0.401 &
0.004\\ V5 & puls & 0.82803* & 15:28:33.09 &-43:13:09.05 & 19.163 &
0.008 & 1.040 & 0.011\\ V27 & puls & 0.53613* & 15:28:43.37
&-43:19:32.32 & 19.792 & 0.011 & 1.311 & 0.014\\ V39 & puls & 0.56404
& 15:28:50.40 &-42:36:22.60 & 19.135 & 0.006 & 1.610 & 0.016\\ V50 &
puls & 0.24001* & 15:29:00.16 &-42:38:27.33 & 17.303 & 0.011 & 0.391 &
0.019\\ V70 & puls & 0.49530 & 15:29:11.40 &-43:19:47.48 & 16.556 &
0.003 & 0.811 & 0.004\\ V71 & puls & 0.76145* & 15:29:13.02
&-42:38:57.31 & 18.734 & 0.007 & 1.842 & 0.008\\ V73 & puls & 1.79230
& 15:29:13.64 &-42:58:51.54 & 17.395 & 0.003 & 0.867 & 0.004\\ V80 &
puls & $\sim$1.23680 & 15:29:16.25 &-43:05:25.57 & 17.635 & 0.003 &
1.287 & 0.004\\ V85 & puls & 1.60200 & 15:29:16.98 &-42:41:11.50 &
18.213 & 0.006 & 1.478 & 0.007\\ V95 & puls & 0.72780 & 15:29:23.21
&-43:14:25.19 & 18.693 & 0.006 & 1.083 & 0.008\\ V100 & puls &
0.27869* & 15:29:26.76 &-43:13:59.07 & 17.250 & 0.012 & 0.469 &
0.016\\ V102 & puls & 0.58834 & 15:29:27.90 &-43:04:49.08 & 19.647 &
0.011 & 1.111 & 0.014\\ V104 & puls & 0.88471 & 15:29:29.65
&-43:14:52.35 & 19.104 & 0.007 & 1.157 & 0.008\\ V124 & puls & 1.50118
& 15:29:43.01 &-42:37:06.05 & 18.490 & 0.006 & 2.185 & 0.007\\ V137 &
puls & 0.86994 & 15:29:46.86 &-42:38:09.34 & 17.432 & 0.004 & 1.322 &
0.005\\ V140 & puls & 1.39000 & 15:29:49.34 &-42:38:15.44 & 16.091 &
0.004 & - & -\\ V162 & puls & 0.87648 & 15:30:02.71
&-42:35:11.54 & 20.541 & 0.022 & 1.454 & 0.026\\ V195 & puls & 0.34534
& 15:30:18.18 &-42:56:41.78 & 18.080 & 0.004 & 1.291 & 0.009\\ V202 &
puls & 0.71040 & 15:30:21.43 &-42:34:09.06 & 20.981 & 0.031 & 2.851 &
0.032\\ V214 & puls & 0.52975 & 15:30:31.73 &-42:33:25.43 & 19.644 &
0.012 & 1.366 & 0.014\\ V217 & puls & 0.24462 & 15:30:32.63
&-42:39:28.83 & 20.391 & 0.016 & 1.480 & 0.019\\ V238 & puls &
0.12377* & 15:30:54.67 &-42:59:57.20 & 18.462 & 0.004 & 0.819 &
0.006\\ V244 & puls & 0.70080 & 15:30:58.27 &-42:44:28.24 & 17.845 &
0.008 & 1.414 & 0.011\\ V249 & puls & 1.78940 & 15:31:00.92
&-42:31:14.49 & 17.138 & 0.003 & 0.779 & 0.004\\ V278 & puls & 1.99370
& 15:31:09.73 &-42:47:30.38 & 17.206 & 0.006 & 0.819 & 0.008\\ V280 &
puls & 1.28910 & 15:31:10.86 &-43:02:24.79 & 18.778 & 0.006 & 0.797 &
0.007\\ V290 & puls & 0.95801 & 15:31:14.65 &-42:32:07.28 & 16.997 &
0.010 & 1.133 & 0.013\\ V291 & puls & 0.26803 & 15:31:15.06
&-42:30:39.15 & 16.279 & 0.003 & 0.333 & 0.004\\ V295 & puls & 0.04159
& 15:31:18.06 &-42:54:01.72 & 17.613 & 0.006 & 0.932 & 0.008\\ V303 &
puls & 1.16566 & 15:31:20.98 &-42:32:34.73 & 19.615 & 0.012 & 1.558 &
0.014\\ V308 & puls & 0.85152 & 15:31:23.53 &-42:50:14.70 & 20.208 &
0.022 & 1.129 & 0.026\\ V313 & puls & 1.51155 & 15:31:24.63
&-43:02:17.25 & 16.690 & 0.003 & 0.731 & 0.006\\ V319 & puls & 0.31448
& 15:31:26.82 &-43:08:55.28 & 18.996 & 0.006 & 0.279 & 0.011\\ V331 &
puls & 1.82460 & 15:31:35.11 &-42:30:24.27 & 18.075 & 0.003 & 0.735 &
0.004\\ V338 & puls & 1.68400 & 15:31:37.62 &-42:36:38.57 & 17.350 &
0.002 & 0.445 & 0.004\\ V353 & puls & 0.91219 & 15:31:45.92
&-42:35:26.45 & 19.262 & 0.008 & 1.081 & 0.010\\ V358 & puls & 0.91176
& 15:31:46.60 &-43:10:18.86 & 18.753 & 0.014 & 0.841 & 0.021\\ V372 &
puls & 0.22507 & 15:31:53.15 &-43:13:25.81 & 20.192 & 0.022 & 3.508 &
0.022\\ V373 & puls & 0.58767 & 15:31:53.69 &-43:06:52.28 & 17.383 &
0.005 & 0.778 & 0.007\\  \hline
\end{tabular}

\centering \footnotesize
\begin{tabular}{lllllllll}
ID&Type&Period (d)&RA (j2000.0)&DEC
(j2000.0)&V&$\sigma$V&V-I&$\sigma$(V-I)\\ & & &
h\hspace{2mm}m\hspace{2mm}s &
$^{\circ}$\hspace{5mm}$'$\hspace{3mm}$''$ & & & & \\
\hline \hline 
V376 & puls & 1.07188 & 15:31:55.27
&-42:32:40.72 & 19.282 & 0.008 & 1.489 & 0.009\\ V391 & puls & $?$ &
15:32:03.91 &-42:51:13.84 & 14.793 & 0.009 & - & -\\ V397 &
puls & 0.87095* & 15:32:06.43 &-42:58:02.41 & 15.720 & 0.003 & 0.602 &
0.005\\ V406 & puls & 1.53755 & 15:32:10.97 &-42:46:01.58 & 18.827 &
0.007 & 1.054 & 0.009\\ V432 & puls & 0.21786 & 15:32:24.34
&-42:31:19.05 & 18.964 & 0.007 & 1.115 & 0.009\\ V434 & puls & 0.21786
& 15:32:25.26 &-42:30:48.94 & 19.226 & 0.016 & 0.700 & 0.023\\ V443 &
puls & 0.62793 & 15:32:29.26 &-43:14:18.54 & 21.011 & 0.029 & 1.567 &
0.033\\ V448 & puls & 1.54900 & 15:32:30.96 &-43:02:02.54 & 18.491 &
0.007 & 1.210 & 0.008\\ V453 & puls & $\sim$0.16550 & 15:32:34.81
&-42:51:04.92 & 17.589 & 0.012 & 1.623 & 0.013\\ V454 & puls & 1.34368
& 15:32:34.91 &-42:47:40.11 & 18.169 & 0.007 & 0.700 & 0.009\\ V489 &
puls & 1.00150 & 15:32:56.86 &-43:00:09.43 & 17.864 & 0.007 & 0.755 &
0.024\\ V491 & puls & $\sim$0.20061 & 15:32:58.17 &-42:52:13.77 &
14.844 & 0.007 & 0.365 & 0.012\\ V1 & LPV & 21.28 & 15:28:29.26 &-43:11:53.40 & 21.035 &
0.057 & 0.993 & 0.069\\ V6 & LPV & 10.30 & 15:28:33.72 &-43:02:05.83 &
16.299 & 0.003 & 1.017 & 0.005\\ V7 & LPV & $\sim$22.73 & 15:28:33.85
&-43:07:08.89 & 17.025 & 0.004 & 0.850 & 0.005\\ V9 & LPV &
$\sim$17.33 & 15:28:34.50 &-43:17:58.58 & 16.618 & 0.007 & 0.911 &
0.009\\ V11 & LPV & 4.0984 & 15:28:36.06 &-42:57:18.28 & 16.732 &
0.003 & 1.037 & 0.005\\ V12 & LPV & $\sim$19.23 & 15:28:37.54
&-43:07:40.62 & 15.984 & 0.004 & 0.686 & 0.007\\ V13 & LPV & 6.875 &
15:28:38.25 &-42:37:39.37 & 15.910 & 0.002 & 0.739 & 0.004\\ V14 & LPV
& 4.01 & 15:28:38.70 &-43:04:08.15 & 17.997 & 0.003 & 0.694 & 0.005\\
V16 & LPV & 2.6865 & 15:28:39.31 &-43:07:00.16 & 18.679 & 0.005 &
0.756 & 0.007\\ V18 & LPV & 6.63 & 15:28:40.72 &-43:03:23.26 & 18.666
& 0.005 & 2.115 & 0.005\\ V19 & LPV & 2.0388 & 15:28:41.04
&-42:32:21.15 & 17.947 & 0.004 & 1.182 & 0.006\\ V21 & LPV & 3.3229 &
15:28:41.54 &-43:13:17.50 & 16.614 & 0.003 & 0.650 & 0.004\\ V22 & LPV
& 2.7833 & 15:28:41.84 &-43:14:16.86 & 18.479 & 0.012 & 1.265 &
0.015\\ V23 & LPV & 14.95 & 15:28:42.35 &-43:02:11.09 & 17.730 & 0.003
& 0.717 & 0.004\\ V25 & LPV & 16.58 & 15:28:42.98 &-42:58:19.01 &
17.476 & 0.004 & 1.087 & 0.004\\ V33 & LPV & 3.923 & 15:28:46.88
&-43:03:26.91 & 18.403 & 0.007 & 1.030 & 0.011\\ V34 & LPV & 12.92 &
15:28:47.61 &-42:38:04.46 & 17.929 & 0.004 & 1.751 & 0.006\\ V35 & LPV
& 7.4 & 15:28:47.71 &-43:19:31.44 & 16.910 & 0.005 & 0.871 & 0.006\\
V38 & LPV & 7.22 & 15:28:50.06 &-43:10:03.62 & 17.598 & 0.003 & 1.141
& 0.004\\ V40 & LPV & 13.87 & 15:28:50.51 &-43:13:16.34 & 16.457 &
0.003 & 0.729 & 0.004\\ V45 & LPV & 7.13 & 15:28:55.18 &-42:37:15.81 &
18.109 & 0.004 & 1.311 & 0.007\\ V47 & LPV & 5.93 & 15:28:57.79
&-43:04:26.36 & 18.242 & 0.005 & 0.840 & 0.007\\ V52 & LPV & 15.72 &
15:29:00.46 &-43:04:46.56 & 18.559 & 0.005 & 1.876 & 0.007\\ V53 & LPV
& $\sim$4.133 & 15:29:00.51 &-42:57:17.27 & 16.041 & 0.002 & 0.696 &
0.004\\ V54 & LPV & 12.61 & 15:29:00.75 &-42:32:19.27 & 16.550 & 0.002
& 0.771 & 0.003\\ V55 & LPV & 4.249 & 15:29:01.43 &-43:02:27.74 &
16.398 & 0.002 & 0.633 & 0.004\\ V56 & LPV & 4.156 & 15:29:01.98
&-43:15:52.40 & 19.618 & 0.009 & 1.253 & 0.011\\ V57 & LPV & 15.10 &
15:29:02.53 &-43:12:12.23 & 15.319 & 0.003 & 0.685 & 0.018\\ V63 & LPV
& $\sim$4.20 & 15:29:06.86 &-43:13:31.45 & 18.098 & 0.031 & 1.352 &
0.039\\ V64 & LPV & $\sim$18.87 & 15:29:08.82 &-43:06:05.73 & 16.803 &
0.003 & 0.754 & 0.005\\ V67 & LPV & 5.37 & 15:29:09.23 &-43:13:16.71 &
19.050 & 0.007 & 1.744 & 0.008\\ V68 & LPV & $\sim$18.38 & 15:29:10.47
&-42:57:44.81 & 17.866 & 0.003 & 1.046 & 0.004\\ V69 & LPV & 5.4065 &
15:29:10.82 &-42:57:15.10 & 18.358 & 0.004 & 1.481 & 0.005\\ V72 & LPV
& $\sim$18.90 & 15:29:13.57 &-43:00:10.78 & 16.256 & 0.002 & 0.870 &
0.004\\ V74 & LPV & 8.197 & 15:29:13.77 &-42:41:26.58 & 18.157 & 0.006
& 0.914 & 0.007\\ V75 & LPV & 3.172 & 15:29:14.49 &-43:07:24.41 &
16.821 & 0.003 & 0.824 & 0.005\\ V77 & LPV & 7.943 & 15:29:14.89
&-42:38:49.80 & 17.904 & 0.003 & 1.108 & 0.005\\ V78 & LPV & 4.19 &
15:29:15.67 &-43:11:06.82 & 17.139 & 0.002 & 0.699 & 0.004\\ \hline
\end{tabular}

\centering \footnotesize
\begin{tabular}{lllllllll}
ID&Type&Period (d)&RA (j2000.0)&DEC
(j2000.0)&V&$\sigma$V&V-I&$\sigma$(V-I)\\ & & &
h\hspace{2mm}m\hspace{2mm}s &
$^{\circ}$\hspace{5mm}$'$\hspace{3mm}$''$ & & & & \\
\hline \hline 
V81 & LPV
& 4.625 & 15:29:16.26 &-42:34:05.06 & 17.227 & 0.003 & 0.826 & 0.004\\
V82 & LPV & 3.580 & 15:29:16.45 &-42:31:40.20 & 17.725 & 0.003 & 1.048
& 0.004\\ V86 & LPV & $\sim$14.70 & 15:29:17.21 &-42:58:44.07 & 17.695
& 0.006 & 1.434 & 0.010\\ V87 & LPV & 14.81 & 15:29:17.49
&-42:57:06.98 & 16.572 & 0.004 & 0.964 & 0.006\\ V88 & LPV &
$\sim$11.24 & 15:29:17.70 &-42:39:15.39 & 16.256 & 0.003 & 0.703 &
0.005\\ V89 & LPV & 8.289 & 15:29:19.68 &-43:05:37.62 & 16.077 & 0.003
& 0.843 & 0.006\\ V90 & LPV & 9.13 & 15:29:20.11 &-43:16:01.66 &
19.108 & 0.008 & 1.367 & 0.009\\ V93 & LPV & 4.543 & 15:29:21.00
&-43:06:22.37 & 19.598 & 0.011 & 1.459 & 0.013\\ V96 & LPV & 16.39 &
15:29:23.49 &-43:01:02.69 & 16.448 & 0.002 & 0.703 & 0.004\\ V97 & LPV
& $\sim$21.28 & 15:29:23.61 &-42:39:09.07 & 17.561 & 0.003 & 0.774 &
0.005\\ V98 & LPV & 5.848 & 15:29:24.95 &-43:07:06.98 & 17.350 & 0.003
& 0.705 & 0.005\\ V99 & LPV & $\sim$21.74 & 15:29:26.05 &-43:18:30.03
& 17.020 & 0.003 & 0.960 & 0.005\\ V101 & LPV & 14.28 & 15:29:27.28
&-42:58:07.87 & 15.906 & 0.002 & 0.814 & 0.010\\ V103 & LPV & 8.816 &
15:29:29.23 &-43:00:59.20 & 17.441 & 0.004 & 0.799 & 0.006\\ V105 &
LPV & 8.98 & 15:29:30.07 &-43:13:26.23 & 17.150 & 0.004 & 0.833 &
0.006\\ V108 & LPV & 8.474 & 15:29:30.89 &-43:14:41.31 & 16.787 &
0.002 & 0.669 & 0.004\\ V109 & LPV & 38.66 & 15:29:34.14 &-42:56:24.65
& 15.247 & 0.004 & - & -\\ V110 & LPV & $\sim$18.87 &
15:29:34.91 &-43:11:31.86 & 15.741 & 0.004 & 1.008 & 0.011\\ V112 &
LPV & 9.584 & 15:29:36.70 &-42:30:02.18 & 16.992 & 0.003 & 0.916 &
0.004\\ V116 & LPV & $\sim$25.00 & 15:29:39.60 &-42:30:09.04 & 15.860
& 0.002 & 0.863 & 0.004\\ V119 & LPV & 20.0 & 15:29:41.62
&-43:06:06.86 & 17.786 & 0.005 & 1.099 & 0.007\\ V120 & LPV & 13.26 &
15:29:41.83 &-43:00:41.58 & 17.357 & 0.002 & 0.741 & 0.005\\ V121 &
LPV & $\sim$10.99 & 15:29:42.00 &-43:20:03.11 & 15.999 & 0.002 & 0.636
& 0.006\\ V122 & LPV & 3.633 & 15:29:42.25 &-43:01:45.55 & 18.306 &
0.006 & 0.892 & 0.008\\ V123 & LPV & 4.635 & 15:29:42.81 &-42:56:28.33
& 17.433 & 0.003 & 1.004 & 0.006\\ V125 & LPV & 7.626 & 15:29:43.30
&-43:05:57.19 & 17.032 & 0.003 & 0.651 & 0.006\\ V126 & LPV & 5.049 &
15:29:43.54 &-43:01:58.36 & 17.495 & 0.003 & 0.865 & 0.006\\ V127 &
LPV & 4.94 & 15:29:43.96 &-43:00:31.74 & 16.126 & 0.002 & 0.770 &
0.008\\ V128 & LPV & 3.765 & 15:29:44.11 &-43:07:20.58 & 17.272 &
0.004 & 0.899 & 0.006\\ V129 & LPV & $\sim$21.74 & 15:29:44.55
&-43:03:22.63 & 15.572 & 0.002 & 0.695 & 0.008\\ V132 & LPV &
$\sim$21.28 & 15:29:45.36 &-42:33:53.01 & 17.222 & 0.002 & 0.734 &
0.004\\ V133 & LPV & $\sim$21.74 & 15:29:45.59 &-42:37:18.88 & 17.024
& 0.004 & 0.977 & 0.006\\ V135 & LPV & 6.90 & 15:29:46.68
&-43:16:19.29 & 18.390 & 0.005 & 0.716 & 0.006\\ V136 & LPV & 23.20 &
15:29:46.69 &-42:40:35.94 & 14.696 & 0.006 & - & -\\ V142 &
LPV & 5.502 & 15:29:49.80 &-42:56:14.31 & 18.756 & 0.004 & 1.130 &
0.008\\ V143 & LPV & 4.58 & 15:29:50.02 &-43:16:55.14 & 18.556 & 0.005
& 0.882 & 0.006\\ V146 & LPV & 5.628 & 15:29:50.61 &-43:00:31.50 & 17.368
& 0.003 & 0.797 & 0.006\\ V147 & LPV & 5.691 & 15:29:51.51
&-43:01:56.35 & 19.497 & 0.012 & 1.108 & 0.016\\ V148 & LPV &
$\sim$9.03 & 15:29:51.74 &-43:16:26.38 & 18.471 & 0.005 & 0.806 &
0.007\\ V150 & LPV & $\sim$18.87 & 15:29:52.21 &-43:18:31.84 & 18.009
& 0.029 & 1.148 & 0.056\\ V152 & LPV & 12.0 & 15:29:57.04
&-43:05:35.49 & 16.840 & 0.003 & 0.865 & 0.007\\ V153 & LPV & 5.76 &
15:29:57.05 &-42:58:32.91 & 15.335 & 0.003 & - & -\\ V154 &
LPV & 24.61 & 15:29:58.15 &-43:17:14.60 & 15.468 & 0.003 & 0.727 &
0.011\\ V156 & LPV & $\sim$7.94 & 15:29:58.56 &-42:33:07.99 & 17.061 &
0.002 & 0.750 & 0.004\\ V157 & LPV & 4.812 & 15:29:59.17 &-43:19:23.88
& 18.543 & 0.005 & 0.916 & 0.008\\ V158 & LPV & 7.411 & 15:29:59.56
&-42:57:25.35 & 18.849 & 0.008 & 0.900 & 0.012\\ V159 & LPV & 6.96 &
15:30:00.52 &-42:59:30.85 & 17.516 & 0.003 & 0.830 & 0.008\\ V160 &
LPV & 6.68 & 15:30:01.77 &-42:39:01.15 & 17.931 & 0.014 & 0.770 &
0.017\\ V163 & LPV & 12.73 & 15:30:03.45 &-42:58:23.94 & 17.173 &
0.003 & 1.109 & 0.009\\ \hline
\end{tabular}

\centering \footnotesize
\begin{tabular}{lllllllll}
ID&Type&Period (d)&RA (j2000.0)&DEC
(j2000.0)&V&$\sigma$V&V-I&$\sigma$(V-I)\\ & & &
h\hspace{2mm}m\hspace{2mm}s &
$^{\circ}$\hspace{5mm}$'$\hspace{3mm}$''$ & & & & \\
\hline \hline 
V164 & LPV & 9.473 & 15:30:03.62 &-42:30:55.12
& 17.905 & 0.004 & 0.956 & 0.006\\ V166 & LPV & 9.76 & 15:30:04.72
&-42:34:57.34 & 18.740 & 0.005 & 1.018 & 0.007\\ V169 & LPV & 3.237 &
15:30:07.93 &-42:31:48.33 & 19.234 & 0.009 & 1.084 & 0.011\\ V173 &
LPV & $\sim$10.2 & 15:30:08.67 &-42:59:36.41 & 15.533 & 0.007 & 0.783
& 0.016\\ V174 & LPV & 7.29 & 15:30:09.08 &-42:38:06.54 & 18.854 &
0.008 & 1.222 & 0.009\\ V175 & LPV & 9.90 & 15:30:09.30 &-43:00:19.69
& 15.727 & 0.003 & 0.695 & 0.016\\ V176 & LPV & $\sim$19.23 &
15:30:10.23 &-42:29:58.96 & 18.020 & 0.007 & 0.981 & 0.009\\ V177 &
LPV & 14.49 & 15:30:11.38 &-43:02:41.60 & 16.159 & 0.004 & 0.747 &
0.010\\ V179 & LPV & 6.93 & 15:30:11.80 &-42:30:54.07 & 17.637 & 0.003
& 0.692 & 0.005\\ V180 & LPV & 9.54 & 15:30:13.09 &-42:33:48.45 &
17.103 & 0.003 & 1.305 & 0.007\\ V182 & LPV & 5.97 & 15:30:13.46
&-42:34:02.34 & 20.133 & 0.012 & 1.893 & 0.028\\ V183 & LPV &
$\sim$22.73 & 15:30:13.52 &-42:38:37.26 & 18.175 & 0.005 & 1.005 &
0.006\\ V185 & LPV & 7.028 & 15:30:15.34 &-43:11:01.75 & 18.177 &
0.005 & 0.722 & 0.008\\ V186 & LPV & $\sim$21.28 & 15:30:15.44
&-43:13:05.93 & 18.175 & 0.009 & 1.190 & 0.010\\ V187 & LPV &
$\sim$18.15 & 15:30:15.86 &-42:38:01.83 & 16.943 & 0.004 & 0.843 &
0.006\\ V190 & LPV & 8.42 & 15:30:17.43 &-43:01:07.30 & 17.023 & 0.004
& 0.799 & 0.007\\ V191 & LPV & 11.65 & 15:30:17.67 &-42:33:39.01 &
20.631 & 0.024 & 1.683 & 0.027\\ V192 & LPV & $\sim$15.63 &
15:30:18.04 &-42:57:17.77 & 15.889 & 0.003 & 0.664 & 0.016\\ V193 &
LPV & 4.372 & 15:30:18.08 &-43:14:18.82 & 17.353 & 0.003 & 0.678 &
0.006\\ V194 & LPV & $\sim$24.39 & 15:30:18.14 &-42:37:14.61 & 17.007
& 0.003 & 0.741 & 0.006\\ V197 & LPV & $\sim$5.051 & 15:30:18.74
&-43:14:15.42 & 16.865 & 0.005 & 0.779 & 0.006\\ V198 & LPV &
$\sim$11.49 & 15:30:19.80 &-43:09:48.32 & 15.683 & 0.003 & 0.569 &
0.009\\ V199 & LPV & 3.533 & 15:30:20.49 &-43:06:34.09 & 17.957 &
0.005 & 1.323 & 0.009\\ V200 & LPV & $\sim$22.22 & 15:30:20.77
&-43:11:51.58 & 16.391 & 0.002 & 0.346 & 0.005\\ V205 & LPV & 2.690 &
15:30:27.36 &-43:11:53.31 & 17.314 & 0.004 & 0.783 & 0.008\\ V208 &
LPV & 7.862 & 15:30:28.89 &-42:57:33.17 & 18.841 & 0.006 & 0.908 &
0.009\\ V209 & LPV & 11.655 & 15:30:29.81 &-42:30:31.79 & 19.182 &
0.005 & 0.938 & 0.008\\ V211 & LPV & $\sim$8.265 & 15:30:31.05
&-43:09:24.79 & 15.285 & 0.005 & - & -\\ V213 & LPV & 8.265
& 15:30:31.49 &-42:58:56.27 & 18.414 & 0.004 & 1.100 & 0.005\\ V215 &
LPV & 8.380 & 15:30:31.99 &-43:17:04.20 & 16.039 & 0.004 & 0.932 &
0.012\\ V216 & LPV & 3.672 & 15:30:32.56 &-43:06:41.07 & 18.964 &
0.016 & 0.907 & 0.024\\ V219 & LPV & 11.90 & 15:30:34.28 &-43:04:34.58
& 16.923 & 0.004 & 0.979 & 0.007\\ V220 & LPV & 16.06 & 15:30:34.54
&-42:56:18.91 & 15.912 & 0.003 & 0.720 & 0.011\\ V222 & LPV & 2.157 &
15:30:34.85 &-42:59:33.80 & 19.619 & 0.011 & 1.429 & 0.013\\ V225 &
LPV & 14.95 & 15:30:36.70 &-43:15:53.27 & 15.418 & 0.006 & 0.780 &
0.022\\ V226 & LPV & 4.604 & 15:30:37.61 &-43:04:28.16 & 18.828 &
0.005 & 1.401 & 0.008\\ V227 & LPV & 5.607 & 15:30:37.78 &-43:13:34.38
& 16.108 & 0.004 & 0.653 & 0.008\\ V229 & LPV & $\sim$19.61 &
15:30:50.56 &-43:15:07.35 & 16.297 & 0.003 & 0.644 & 0.008\\ V230 &
LPV & 5.10 & 15:30:50.96 &-42:50:31.11 & 17.195 & 0.008 & 0.778 &
0.010\\ V233 & LPV & 4.344 & 15:30:52.44 &-43:15:41.16 & 19.401 &
0.011 & 1.162 & 0.012\\ V234 & LPV & $\sim$21.74 & 15:30:52.50
&-43:20:26.77 & 17.889 & 0.004 & 0.827 & 0.006\\ V235 & LPV & 10.74 &
15:30:53.22 &-43:17:34.47 & 18.600 & 0.005 & 0.799 & 0.009\\ V237 &
LPV & 9.60 & 15:30:54.29 &-42:39:09.08 & 16.098 & 0.003 & 0.662 &
0.006\\ V239 & LPV & 18.61 & 15:30:55.37 &-43:06:19.67 & 17.466 &
0.015 & 0.651 & 0.025\\ V240 & LPV & 5.44 & 15:30:55.44 &-42:58:06.47
& 17.779 & 0.004 & 0.767 & 0.006\\ V242 & LPV & 8.064 & 15:30:57.30
&-43:07:23.12 & 18.372 & 0.006 & 1.060 & 0.009\\ V245 & LPV &
$\sim$5.076 & 15:30:59.23 &-43:19:50.79 & 17.262 & 0.006 & 1.535 &
0.008\\
\hline
\end{tabular}

\centering \footnotesize
\begin{tabular}{lllllllll}
ID&Type&Period (d)&RA (j2000.0)&DEC
(j2000.0)&V&$\sigma$V&V-I&$\sigma$(V-I)\\ & & &
h\hspace{2mm}m\hspace{2mm}s &
$^{\circ}$\hspace{5mm}$'$\hspace{3mm}$''$ & & & & \\
\hline \hline 
 V255 & LPV & 11.62 & 15:31:02.34 &-42:39:19.19 & 19.645 &
0.009 & 1.782 & 0.011\\ V256 & LPV & 8.279 & 15:31:02.85 &-42:38:57.92
& 18.181 & 0.009 & 0.956 & 0.019\\ 
V258 & LPV & 4.244 & 15:31:03.08 &-42:34:49.00 & 19.364 & 0.008 & 1.034 & 0.011\\ V259 & LPV &
$\sim$2.02 & 15:31:03.33 &-42:53:23.32 & - & - & - &
-\\ V260 & LPV & $\sim$23.26 & 15:31:03.83 &-42:30:32.35 & 17.409
& 0.003 & 1.110 & 0.005\\ V263 & LPV & 10.68 & 15:31:05.18
&-42:30:47.20 & 18.857 & 0.006 & 2.339 & 0.007\\ V265 & LPV & 17.58 &
15:31:05.57 &-42:40:51.21 & 18.233 & 0.027 & 0.854 & 0.037\\ V266 &
LPV & $\sim$10.87 & 15:31:05.82 &-43:08:50.85 & 16.353 & 0.003 & 0.937
& 0.005\\ V267 & LPV & $\sim$18.18 & 15:31:06.04 &-43:14:49.65 &
16.703 & 0.003 & 0.787 & 0.007\\ V269 & LPV & $\sim$15.15 &
15:31:06.46 &-42:36:56.38 & 15.946 & 0.002 & 0.844 & 0.006\\ V273 &
LPV & 3.76 & 15:31:08.18 &-43:14:33.15 & 19.256 & 0.007 & 1.008 &
0.010\\ V274 & LPV & 13.39 & 15:31:08.60 &-43:04:49.05 & 17.329 &
0.004 & 0.735 & 0.006\\ V275 & LPV & 9.140 & 15:31:08.82 &-43:00:31.92
& 18.403 & 0.010 & 1.259 & 0.011\\ V279 & LPV & $\sim$7.430 &
15:31:10.25 &-42:55:51.91 & 16.386 & 0.005 & 0.724 & 0.006\\ 
V282 & LPV & 8.850 & 15:31:11.42 &-42:52:51.22 & 16.956
& 0.005 & 0.860 & 0.007\\ V286 & LPV & 7.403 & 15:31:13.34
&-42:58:31.08 & 18.954 & 0.006 & 0.761 & 0.008\\ V287 & LPV & 15.11 &
15:31:13.68 &-43:12:27.67 & 18.527 & 0.005 & 1.438 & 0.006\\ V289 &
LPV & 7.239 & 15:31:14.27 &-42:31:59.50 & 18.160 & 0.005 & 1.084 &
0.006\\ V292 & LPV & 4.696 & 15:31:16.76 &-43:17:04.43 & 15.689 &
0.005 & 0.660 & 0.011\\ V298 & LPV & 2.495 & 15:31:19.73 &-42:33:30.86
& 19.122 & 0.012 & 1.644 & 0.019\\ V299 & LPV & 9.023 & 15:31:19.74
&-42:57:18.36 & 18.225 & 0.005 & 0.777 & 0.007\\ V300 & LPV & 7.370 &
15:31:20.32 &-42:52:18.13 & 16.084 & 0.005 & 0.719 & 0.006\\ V301 &
LPV & $\sim$18.519 & 15:31:20.38 &-43:13:45.95 & 18.989 & 0.007 &
0.745 & 0.011\\ V309 & LPV & 2.036 & 15:31:23.75 &-43:12:25.97 &
17.868 & 0.004 & 1.151 & 0.006\\ V311 & LPV & 16.05 & 15:31:23.83
&-42:44:30.76 & 17.578 & 0.006 & 1.036 & 0.009\\ V314 & LPV & 6.029 &
15:31:24.82 &-42:40:01.30 & 17.417 & 0.004 & 0.977 & 0.006\\ V315 &
LPV & 4.024 & 15:31:25.00 &-42:51:32.36 & 16.598 & 0.005 & 0.809 &
0.006\\ V316 & LPV & 16.10 & 15:31:25.19 &-42:47:06.86 & 18.696 &
0.026 & 1.193 & 0.043\\ V318 & LPV & 2.746 & 15:31:26.16 &-42:29:50.65
& 16.776 & 0.002 & 0.851 & 0.004\\ V320 & LPV & $\sim$24.39 &
15:31:27.01 &-42:36:50.50 & 18.609 & 0.005 & 1.981 & 0.006\\ V321 &
LPV & 7.79 & 15:31:27.47 &-42:50:27.26 & 17.666 & 0.006 & 0.860 &
0.009\\ V323 & LPV & 4.34 & 15:31:29.14 &-43:19:55.83 & 18.399 & 0.005
& 0.839 & 0.007\\ V325 & LPV & 6.917 & 15:31:30.25 &-42:39:06.38 &
16.246 & 0.054 & 0.953 & 0.054\\ V326 & LPV & 6.79 & 15:31:30.76
&-43:16:53.56 & 17.712 & 0.008 & 0.644 & 0.011\\ V327 & LPV & 6.81 &
15:31:30.86 &-43:16:55.69 & 18.641 & 0.021 & 0.566 & 0.028\\ V328 &
LPV & 3.853 & 15:31:31.35 &-42:33:20.86 & 19.059 & 0.006 & 1.015 &
0.009\\ V329 & LPV & 9.58 & 15:31:33.46 &-43:19:23.48 & 19.135 & 0.011
& 0.585 & 0.014\\ V332 & LPV & 23.20 & 15:31:35.22 &-43:12:28.03 &
20.490 & 0.072 & 0.558 & 0.088\\ V335 & LPV & 3.2955 & 15:31:37.20
&-42:43:50.62 & 17.170 & 0.007 & 0.826 & 0.015\\ V341 & LPV &
$\sim$23.26 & 15:31:38.47 &-42:50:15.44 & 16.249 & 0.003 & 0.812 &
0.005\\ V343 & LPV & 3.647 & 15:31:38.97 &-42:29:36.15 & 17.722 &
0.004 & 1.430 & 0.005\\ V346 & LPV & 6.329 & 15:31:42.00 &-42:32:28.47
& 18.908 & 0.006 & 0.968 & 0.009\\ V348 & LPV & 2.202 & 15:31:43.03
&-43:09:49.22 & 18.350 & 0.003 & 1.075 & 0.006\\ V349 & LPV & 2.310 &
15:31:43.12 &-42:37:34.64 & 18.349 & 0.005 & 1.603 & 0.006\\ V350 &
LPV & $\sim$21.74 & 15:31:43.55 &-43:15:40.75 & 17.865 & 0.004 & 0.649
& 0.008\\ V357 & LPV & 13.19 & 15:31:46.52 &-42:30:38.86 & 18.336 &
0.007 & 0.768 & 0.011\\ V360 & LPV & 13.90 & 15:31:47.62 &-43:12:18.67
& 17.943 & 0.017 & 1.268 & 0.026\\ V361 & LPV & 21.75 & 15:31:47.73
&-43:12:57.97 & 16.451 & 0.002 & 0.760 & 0.006\\ V362 & LPV & 17.31 &
15:31:48.88 &-43:13:58.33 & 15.969 & 0.002 & 0.653 & 0.004\\ 
\hline
\end{tabular}

\centering \footnotesize
\begin{tabular}{lllllllll}
ID&Type&Period (d)&RA (j2000.0)&DEC
(j2000.0)&V&$\sigma$V&V-I&$\sigma$(V-I)\\ & & &
h\hspace{2mm}m\hspace{2mm}s &
$^{\circ}$\hspace{5mm}$'$\hspace{3mm}$''$ & & & & \\
\hline \hline 
V364 &
LPV & 7.183 & 15:31:50.64 &-43:15:33.98 & 19.242 & 0.013 & 2.031 &
0.014\\ V370 & LPV & 3.916 & 15:31:52.09 &-42:51:34.77 & 17.090 &
0.007 & 0.815 & 0.009\\ V375 & LPV & 5.916 & 15:31:54.82 &-42:36:09.20
& 19.327 & 0.010 & 1.665 & 0.011\\ V377 & LPV & 8.282 & 15:31:55.78
&-43:02:13.09 & 18.291 & 0.004 & 0.884 & 0.006\\ V379 & LPV & 31.00 &
15:31:56.58 &-43:06:35.71 & 14.847 & 0.008 & - & -\\ V380 &
LPV & 4.431 & 15:31:56.78 &-43:00:03.54 & 19.757 & 0.012 & 1.131 &
0.014\\ V382 & LPV & 21.74 & 15:31:57.59 &-42:53:10.59 & 14.406 &
0.012 & - & -\\ V384 & LPV & 7.911 & 15:31:58.55
&-43:04:09.71 & 19.561 & 0.012 & 1.228 & 0.014\\ V386 & LPV & 4.922 &
15:31:59.54 &-43:04:16.24 & 16.826 & 0.017 & 0.769 & 0.023\\ V388 &
LPV & 6.328 & 15:32:00.25 &-43:16:32.92 & 18.569 & 0.005 & 0.817 &
0.006\\ V389 & LPV & 6.577 & 15:32:03.19 &-43:15:36.28 & 19.218 &
0.008 & 0.846 & 0.014\\ V390 & LPV & 9.843 & 15:32:03.59 &-42:31:57.77
& 17.897 & 0.003 & 0.832 & 0.005\\ V392 & LPV & 14.39 & 15:32:03.92
&-42:45:57.27 & 17.061 & 0.005 & 1.184 & 0.006\\ V394 & LPV & 14.11 &
15:32:05.53 &-42:59:39.29 & 18.574 & 0.025 & 1.658 & 0.027\\ V396 &
LPV & 6.996 & 15:32:06.29 &-43:17:28.12 & 15.664 & 0.001 & -1.849 &
0.016\\ V398 & LPV & 6.602 & 15:32:06.52 &-43:17:30.54 & 18.163 &
0.021 & 0.650 & 0.026\\ V401 & LPV & $\sim$9.08 & 15:32:07.63
&-43:15:44.87 & 17.677 & 0.003 & 0.308 & 0.005\\ V407 & LPV & 3.398 &
15:32:11.18 &-42:32:18.10 & 17.209 & 0.002 & 1.314 & 0.004\\ V408 &
LPV & 9.35 & 15:32:11.25 &-43:19:17.71 & 16.726 & 0.003 & 0.759 &
0.004\\ V409 & LPV & 10.95 & 15:32:12.28 &-42:31:43.23 & 14.979 &
0.004 & - & -\\ V412 & LPV & $\sim$2.364 & 15:32:15.43
&-42:59:21.73 & 18.557 & 0.008 & 1.201 & 0.011\\ V414 & LPV &
$\sim$3.984 & 15:32:16.05 &-42:53:12.41 & 18.563 & 0.006 & 1.134 &
0.008\\ V415 & LPV & 18.89 & 15:32:17.39 &-43:16:06.34 & 16.825 &
0.003 & 1.002 & 0.006\\ V417 & LPV & 9.12 & 15:32:18.21 &-42:53:09.06
& 19.535 & 0.013 & 2.234 & 0.014\\ V420 & LPV & 7.307 & 15:32:19.20
&-42:36:59.71 & 17.591 & 0.005 & 0.778 & 0.006\\ V421 & LPV & 2.4611 &
15:32:19.45 &-43:11:46.24 & 17.894 & 0.003 & 0.693 & 0.005\\ V423 &
LPV & 15.395 & 15:32:20.27 &-42:42:53.41 & 18.853 & 0.008 & 1.077 &
0.014\\ V425 & LPV & 5.799 & 15:32:21.73 &-43:09:32.51 & 17.722 &
0.003 & 0.704 & 0.006\\ V427 & LPV & 10.04 & 15:32:23.17 &-42:37:03.14
& 18.129 & 0.011 & 1.120 & 0.012\\ V436 & LPV & 7.40 & 15:32:25.57
&-43:12:05.03 & 17.490 & 0.002 & 0.790 & 0.004\\ V439 & LPV & 9.78 &
15:32:28.64 &-43:01:31.67 & 16.585 & 0.002 & 0.669 & 0.004\\ V440 &
LPV & 6.30 & 15:32:28.80 &-42:57:02.91 & 17.978 & 0.004 & 0.898 &
0.006\\ V447 & LPV & 7.319 & 15:32:30.81 &-42:39:33.69 & 19.734 &
0.013 & 1.127 & 0.016\\ V450 & LPV & $\sim$4.88 & 15:32:32.80
&-43:18:14.46 & 18.106 & 0.004 & 2.023 & 0.006\\ V455 & LPV & 8.665 &
15:32:35.39 &-43:02:01.63 & 18.213 & 0.004 & 1.000 & 0.006\\ V460 &
LPV & 21.40 & 15:32:36.69 &-42:56:21.76 & 18.761 & 0.007 & 2.719 &
0.008\\ V463 & LPV & 6.551 & 15:32:37.37 &-43:09:54.35 & 19.994 &
0.014 & 1.069 & 0.017\\ V465 & LPV & 10.22 & 15:32:39.58 &-42:39:37.19
& 17.823 & 0.003 & 1.069 & 0.005\\ V466 & LPV & 10.66 & 15:32:39.60
&-42:48:02.25 & 18.194 & 0.016 & 0.969 & 0.022\\ 
V468 & LPV & 10.32 & 15:32:40.96 &-42:31:23.08 & 20.099
& 0.015 & 1.466 & 0.017\\ V470 & LPV & 5.01 & 15:32:42.84
&-43:06:04.83 & 17.263 & 0.004 & 0.616 & 0.007\\ V471 & LPV & 8.92 &
15:32:45.15 &-43:18:08.79 & 17.662 & 0.005 & 0.749 & 0.008\\ V473 &
LPV & 4.80 & 15:32:45.51 &-43:16:44.77 & 17.343 & 0.004 & 0.998 &
0.006\\ V474 & LPV & 12.60 & 15:32:45.68 &-42:47:48.89 & 15.892 &
0.004 & 0.825 & 0.006\\ V475 & LPV & 9.96 & 15:32:46.57 &-43:03:30.21
& 14.799 & 0.010 & - & -\\ V478 & LPV & $\sim$21.28 &
15:32:48.17 &-42:31:57.88 & 18.660 & 0.005 & 1.312 & 0.007\\ V479 &
LPV & 13.38 & 15:32:48.49 &-42:52:14.25 & 18.404 & 0.006 & 0.916 &
0.008\\ 
\hline
\end{tabular}

\centering \footnotesize
\begin{tabular}{lllllllll}
ID&Type&Period (d)&RA (j2000.0)&DEC
(j2000.0)&V&$\sigma$V&V-I&$\sigma$(V-I)\\ & & &
h\hspace{2mm}m\hspace{2mm}s &
$^{\circ}$\hspace{5mm}$'$\hspace{3mm}$''$ & & & & \\
\hline \hline 
V484 & LPV & $\sim$22.22 & 15:32:52.17 &-43:14:19.93 & 14.757
& 0.010 & - & -\\ V485 & LPV & 4.984 & 15:32:53.44
&-42:42:25.61 & 18.586 & 0.006 & 1.646 & 0.008\\ V486 & LPV & 6.44 &
15:32:55.40 &-43:04:04.95 & 17.671 & 0.004 & 0.699 & 0.007\\ V487 &
LPV & 12.50 & 15:32:55.49 &-42:44:05.71 & 16.495 & 0.006 & 1.156 &
0.009\\ V488 & LPV & $\sim$23.26 & 15:32:55.98 &-43:05:08.96 & 16.850
& 0.003 & 1.050 & 0.007\\ V490 & LPV & 9.77 & 15:32:57.22
&-43:16:03.09 & 17.690 & 0.005 & 0.697 & 0.009\\ V493 & LPV & 5.279 &
15:32:58.45 &-43:02:20.01 & 15.956 & 0.002 & 0.615 & 0.006\\ V134 & Irr & ?  & 15:29:45.87 &-42:55:56.57 & 16.647 & 0.003
& 0.349 & 0.006\\ \hline
\end{tabular}

\vspace*{3mm} \normalsize{Table 2: Table of all detected variable
stars in our Lupus field. Tabulated are the variable star
identification number, the type of variable, the period, RA and DEC
(given in J2000.0), V magnitude, V error, V$-$I color and its
associated uncertainty. Those pulsators with periods marked as $\ast$
are those which could be eclipsing binaries with twice the tabulated
period. Unknown magnitude values are marked as `$-$'}


\begin{thebibliography}{}
\bibitem[Alard \& Lupton(1998)]{AL98} Alard, C.~\& Lupton, R.~H.\
1998, \apj, 503, 325
\bibitem[Albrow et al.(2001)]{A2001} Albrow, M.~D., Gilliland, R.~L.,
Brown, T.~M., Edmonds, P.~D., Guhathakurta, P., \& Sarajedini, A.\
2001, \apj, 559, 1060
\bibitem[Alcock et al.(2003)]{Alc2003} Alcock, C., et al.\ 2003, \apj,
598, 597
\bibitem[Bayliss \& Sackett(2007)]{BS2007} Bayliss, D.~D.~R., \&
Sackett, P.~D.\ 2007, Transiting Extrasolar Planets Workshop, 366, 320
\bibitem[Blazhko (1907)]{Bla07} Blazhko, S. \ 1907, Astron. Nachr.,
175, 325
\bibitem[Bretthorst(2001)]{Brett01} Bretthorst, G.~L.\ 2001, AIP
Conf.~Proc.~568: Bayesian Inference and Maximum Entropy Methods in
Science and Engineering, 568, 241
\bibitem[Cacciari \& Clementini(2003)]{Ca2003} Cacciari, C., \&
Clementini, G.\ 2003, LNP Vol.~635: Stellar Candles for the
Extragalactic Distance Scale, 635, 105
\bibitem[Clementini et al.(2003)]{Cl2003} Clementini, G., Gratton, R.,
Bragaglia, A., Carretta, E., Di Fabrizio, L., \& Maio, M.\ 2003, \aj,
125, 1309
\bibitem[Deeg et al.(2004)]{Deeg2004} Deeg, H.~J., Alonso, R.,
Belmonte, J.~A., Alsubai, K., Horne, K., \& Doyle, L.\ 2004, \pasp,
116, 985
\bibitem[Gratton et al.(2004)]{G2004} Gratton, R.~G., Bragaglia, A.,
Clementini, G., Carretta, E., Di Fabrizio, L., Maio, M., \& Taribello,
E.\ 2004, \aap, 421, 937
\bibitem[Hartman et al.(2004)]{H2004} Hartman, J.~D., Bakos, G.,
Stanek, K.~Z., \& Noyes, R.~W.\ 2004, \aj, 128, 1761
\bibitem[Kane et al.(2005)]{Kane2005} Kane, S.~R., Lister, T.~A.,
Collier Cameron, A., Horne, K., James, D., Pollacco, D.~L., Street,
R.~A., \& Tsapras, Y.\ 2005, \mnras, 362, 117
\bibitem[Landolt(1992)]{Lan1992} Landolt, A.~U.\ 1992, \aj, 104, 340
\bibitem[Pepper \& Burke(2006)]{PB2006} Pepper, J., \& Burke, C.~J.\
2006, \aj, 132, 1177
\bibitem[Rich et al.(2005)]{R2005} Rich, R.~M., Corsi, C.~E.,
Cacciari, C., Federici, L., Fusi Pecci, F., Djorgovski, S.~G., \&
Freedman, W.~L.\ 2005, \aj, 129, 2670
\bibitem[Rucinski(2007)]{Ruc2007} Rucinski, S. \ 2007, ArXiv
Astrophysics e-prints, arXiv:astro-ph/0708.3020
\bibitem[Sandage(1981a)]{Sand1981a} Sandage, A.\ 1981, \apjl, 244, L23
\bibitem[Sandage(1981b)]{Sand1981b} Sandage, A.\ 1981, \apj, 248, 161
\bibitem[Sandage, Diethelm, \& Tammann(1994)]{Sand94} Sandage, A.,
Diethelm, R., \& Tammann, G.~A.\ 1994, \aap, 283, 111
\bibitem[Schlegel et al.(1998)]{Sch1998} Schlegel, D.~J., Finkbeiner,
D.~P., \& Davis, M.\ 1998, \apj, 500, 525
\bibitem[Schwarzenberg-Czerny(1989)]{S1989} Schwarzenberg-Czerny, A.\
1989, \mnras, 241, 153
\bibitem[Soszy{\'n}ski(2006)]{Sos2006} Soszy{\'n}ski, I.\ 2006,
Memorie della Societa Astronomica Italiana, 77, 265
\bibitem[Stetson(2000)]{Stet2000} Stetson, P.~B.\ 2000, \pasp, 112,
925
\bibitem[Tamuz et al.(2005)]{T2005} Tamuz, O., Mazeh, T., \& Zucker,
S.\ 2005, \mnras, 356, 1466
\bibitem[Vivas et al.(2001)]{Vivas01} Vivas, A.~K., et al.\ 2001,
\apjl, 554, L33
\bibitem[Weldrake et al.(2004)]{W2004} Weldrake, D.~T.~F., Sackett,
P.~D., Bridges, T.~J., \& Freeman, K.~C.\ 2004, \aj, 128, 736
\bibitem[Weldrake et al.(2005)]{W2005} Weldrake, D.~T.~F., Sackett,
P.~D., Bridges, T.~J., \& Freeman, K.~C.\ 2005, \apj, 620, 1043
\bibitem[Weldrake et al.(2006)]{W2006} Weldrake, D.~T.~F, Sackett,
P.~D, \& Bridges, T.~J 2006, ArXiv Astrophysics e-prints,
arXiv:astro-ph/0612215
\bibitem[Weldrake et al.(2007a)]{W2007a} Weldrake, D.~T.~F., Sackett,
P.~D., \& Bridges, T.~J.\ 2007a, \aj, 133, 1447
\bibitem[Weldrake et al.(2007c)]{W2007c} Weldrake, D.~T., 
Sackett, P.~D, Bridges, T.~J, \& .\ 2007c, ArXiv e-prints, 710, 
arXiv:0710.3461 
\bibitem[Weldrake et al.(2007b)]{W2007b} Weldrake, D.~T.~F., 
Bayliss, D.~D.~R., Sackett, P.~D., Tingley, B.~W., Gillon, M., \& Setiawan, 
J.\ 2007b, ArXiv e-prints, 711, arXiv:0711.1746 
\bibitem[Wo{\'z}niak et al.(2004)]{Woz2004} Wo{\'z}niak, P.~R., et
al.\ 2004, \aj, 127, 2436
\bibitem[Wozniak(2000)]{Woz2000} Wozniak, P.~R.\ 2000, Acta
Astronomica, 50, 421
\bibitem[Zacharias et al.(2004)]{Zach2004} Zacharias, N., Monet,
D.~G., Levine, S.~E., Urban, S.~E., Gaume, R., \& Wycoff, G.~L.\ 2004,
Bulletin of the American Astronomical Society, 36, 1418
\end{thebibliography}
\end{document}